# Numerical study of the behaviors of ventilated supercavities in a periodic gust flow

Renfang Huang[1,2], Siyao Shao[2,3], Roger E. A. Arndt[2], Xianwu Luo[4], Jiarong Hong[2,3 *]

1. Key Laboratory for Mechanics in Fluid Solid Coupling Systems, Institute of Mechanics, Chinese Academy of Sciences, Beijing, China 100190.
2. Saint Anthony Falls Laboratory, 2 3rd AVE SE, University of Minnesota, Minneapolis, MN, USA 55414.
3. Department of Mechanical Engineering, University of Minnesota, Minneapolis, MN, USA 55414.
4. Department of Energy and Power Engineering, Tsinghua University, Beijing China 100084.

* Email addresses of the corresponding author: jhong@umn.edu

## Abstract

We conducted a numerical simulation of ventilated supercavitation from a forward-facing cavitator in unsteady flows generated by a gust generator under different gust angles of attack and gust frequencies. The numerical method is validated through the experimental results under specific steady and unsteady conditions. It has been shown that the simulation can capture the degree of cavity shape fluctuation and internal pressure variation in a gust cycle. Specifically, the cavity centerline shows periodic wavelike undulation with a maximum amplitude matching that of the incoming flow perturbation. The cavity internal pressure also fluctuates periodically, causing the corresponding change of difference between internal and external pressure across the closure that leads to the closure mode change in a gust cycle. In addition, the simulation captures the variation of cavity internal flow, particularly the development internal flow boundary layer along the cavitator mounting strut, upon the incoming flow perturbation, correlating with cavity deformation and closure mode variation. With increasing angle of attack, the cavity exhibits augmented wavelike undulation and pressure fluctuation. As the wavelength of the flow perturbation approaches the cavity length with increasing gust frequency, the cavity experiences stronger wavelike undulation and internal pressure fluctuation but reduced cavitation number variation.

## 1. Introduction

Supercavitation is a special case of cavitation in which a cavity is large enough to encompass the object travelling in the liquid (Franc and Michel 2006, Nesteruk 2012). It provides a promising approach to augment speed of underwater vehicles with drag reduced by as much as 90% (Ceccio 2010). A supercavity can be formed at low speeds by ventilating non-condensable gas into the low-pressure region around a cavitating object (i.e., cavitator), termed as artificial or ventilated supercavitation (Kawakami and Arndt 2011). Ventilated supercavitation is generally characterized by using several non-dimensional parameters, including ventilated cavitation number $\sigma = 2(p_\infty - p)/(\rho_w U_\infty^2)$, Froude number $Fr = U_\infty/\sqrt{gd_c}$ and air entrainment coefficient $C_Q = \dot{Q}/U_\infty d_c^2$, where $p_\infty$ is the ambient pressure upstream of the cavitator, $p_c$ is the cavity pressure measured inside the cavity, $\rho_w$, $U_\infty$, $g$ refer to the water density, the upstream incoming flow velocity at the test-section and gravitational acceleration, respectively, $d_c$ corresponds to the cavitator diameter and $\dot{Q}$ denotes the volumetric ventilation rate. A number of researchers have studied the ventilated supercavitation focusing on the mechanism of the gas leakage and the ventilation demand to form and sustain the supercavity (Savchenko 2001). However, there is no consensus on the analytical models for the supercavity closure and the gas ventilation (Semenenko 2001), and no consensus on the formation conditions for different closure modes (Kawakami and Arndt 2011, Karn et al. 2016). Karn et al. (2016) investigated systematically the closure modes of ventilated supercavity under different flow and ventilation conditions using a backward-facing cavitator model at Saint Anthony Falls Laboratory (SAFL). Their study not only revealed several new closure modes, but also provided a unified physical framework to explain the variation and transition of different closure modes. Using particle image velocimetry, a recent investigation (Wu et al. 2019) studied the internal flow of a ventilated supercavity under different closure types. The internal flow results suggest that at the upstream of the location of the maximum cavity diameter, the gas enters the forward flow (including the internal boundary layer and the forward moving portion of the ventilation influence region) from the reverse flow, while at the downstream of that location, the gas is stripped from the internal boundary layer and enters the reverse flow due to the increasing adverse pressure gradient in the streamwise direction. Combined with visualization results of cavity geometry and closure patterns, the above results are able to explain the influence of the cavity gas leakage mechanisms on cavity growth and closure transition. However, due to immerse technical difficulties, the experimental study of cavity internal flow has yet calculated the pressure variation inside the cavity and the cavity internal flow under unsteady flow conditions has not been conducted.

In practical applications, the supercavitating vehicle is expected to be operated under unsteady external flow conditions. Particularly, when the underwater vehicle travels near the sea surface, the supercavity may encounter an unsteady incoming flow induced by the sea waves. However, there is only a very limited number of experimental works on supercavitation in unsteady flows. For example, Lee et al. (2013) experimentally studied the ventilated supercavitation around a vehicle pitching up and down with the emphasis on the interaction between the vehicle aft body and supercavity boundary. Lee et al. (2013) duplicated several sea states using a periodic gust generator in the high-speed cavitation tunnel at SAFL to investigate the effects of unsteady flows on the shape deformation of axisymmetric supercavities. Using the same setup, Karn et al. (2015) provided further insight into the dependence of supercavity closure on the flow unsteadiness, and found the incoming unsteady flows would cause the transition of the supercavity closure and its shape. Systematic experiments were performed by Shao et al. (2018) to explore different states of a ventilated supercavity under various unsteady flow settings. Although synchronized measurements of cavity shape and pressure are conducted in the past (Lee et al. 2013, Karn et al. 2015, and Shao et al. 2018), the pressure inside the ventilated supercavity is measured at a fixed location and the information regarding to the internal velocity and pressure field is still quite scarce in periodic gust flows due to the limitation of diagnostic approaches.

With the advancement of computational fluid dynamics, numerical simulation has been implemented to provide a more detailed understanding of the cavity internal flows and the characteristics of supercavitation under the conditions which cannot be readily achieved by the experiments. Specifically, using Reynolds-averaged-Navier-Stokes (RANS) simulation, Cao et al. (2017) investigated the pressure distribution inside a ventilated supercavity under different ventilation rates and the blockages in a closed-wall water tunnel. The work highlighted the strong pressure variation near the closure region of the ventilated supercavity, which was potentially connected with different closure modes as pointed out in Karn et al. (2016). Rashidi et al. (2014) used the VOF method coupling with Youngs' algorithm in the advection of the free-surface to investigate physics of ventilated cavitation phenomena (i.e. the cavity shape, the gas leakage and the re-entrant jet), and the re-entrant jet was found to cause the variations of the cavity pressure and the transient flow behaviors such as the cavity detachment and the internal flow inside the cavity. Using a free surface model and a filter-based approach, Wang et al. (2015) numerically investigated the gas leakage behavior and re-entrant jet dynamics for a vehicle body from partial cavitation to supercavitation. They found two mechanisms of the gas leakage at different Froude numbers, termed toroidal vortices mode and two hollow tube vortices mode, and the re-entrant jet at the cavity tail was shown due to the re-circulation of water into the cavity. By using the SST $k$-$\omega$ turbulence model with a mass transfer modeling, the natural cavitation and supercavitation were studied around submarine hull shapes with/without sail and appendages Shang (2013). The sail and appendages were found to promote the occurrence of the supercavity. Park and Rhee (2012) used the full cavitation model with the van Leer scheme to simulate the natural supercavity around a two-dimensional wedge-shaped cavitator for various wedge angle and cavitation numbers, and the results were in good agreement with those obtained with an analytic solution and the potential flow solver. However, effects of the incoming flow unsteadiness on the internal flow and pressure fields inside the supercavity have not yet been numerically explored and investigated in detail.

The current paper presents some numerical results for the ventilated supercavity in a periodic gust flow with different fluctuating amplitudes and frequencies, particularly focusing on the dependence of the supercavity deformation and internal field variations upon the periodic gust flow. The paper is structured as follows: section 2 provides a detailed description of the numerical approach including the governing equations and the simulation setup as well as the experiment for validating the simulation. Subsequently, the section 3 presents the results of the supercavity deformation and transient internal flows under various unsteady flow settings. Finally, a summary and discussion of the results is provided in section 4.

## 2. Research Approach

### 2.1 Governing equations and computational setup

The inhomogeneous hydrodynamic equations under the Eulerian framework are used to formulate the water/air two-phase flows where both water and air are treated as continuous fluids. The various fluid components are assumed to have the interfacial velocity slip and share the same pressure with the interphase mass transfer neglected due to the present high ambient pressure. The RANS equations for the unsteady three-dimensional incompressible flows are as the following:

Continuity equations: $$\frac{\partial}{\partial t}(\alpha_k \rho_k) + \nabla \cdot (\alpha_k \rho_k \mathbf{u}_k) = 0 \tag{1}$$

Volume fraction equation: $$\sum \alpha_k = 1 \tag{2}$$

Where $\alpha_k$, $\rho_k$ presents the volume fraction and density of phase $k$ , respectively, and $\mathbf{u}_k$ is the mean velocity vector. The subscript $k$ denotes the phase water if $k = w$ and the phase air if $k = a$. The water density is $\rho_{\mathrm{w}} = 997\ \mathrm{kg/m^3}$ and the air density is $\rho_{\mathrm{a}} = 1.185\ \mathrm{kg/m^3}$ .

Momentum equations: $$\frac{\partial}{\partial t}(\alpha_k \rho_k \mathbf{u}_k) + \nabla \cdot (\alpha_k \rho_k \mathbf{u}_k \mathbf{u}_k) = -\alpha_k \nabla p + \nabla \cdot \left(\alpha_k \left(\mu_k + \mu_{t,k}\right)(\nabla \mathbf{u}_k + (\nabla \mathbf{u}_k)^{\mathrm{T}})\right) + \mathbf{S}_{k,\mathrm{buoy}} + \mathbf{M}_k \tag{3}$$

where $p$ is the pressure, $\mu_k$is the molecular dynamic viscosity of phase $k$ and $\mu_{t,k}$ corresponds to turbulent eddy viscosity of phase $k$ which is given by the standard $k - \varepsilon$ turbulence model (Launder and Spalding 1974). $\mathbf{S}_{k,\mathrm{buoy}} = (\rho_{\mathrm{a}} - \rho_{\mathrm{w}})\mathbf{g}$ corresponds to the momentum source due to buoyancy forces. $\mathbf{M}_k = C_{\mathrm{D}} \rho_{\mathrm{m}} A_{\mathrm{m}} |\mathbf{u}_{\mathrm{a}} - \mathbf{u}_{\mathrm{w}}| (\mathbf{u}_{\mathrm{a}} - \mathbf{u}_{\mathrm{w}})$ is the total interfacial drag forces acting on the phase water due to the presence of phase air, where the non-dimensional drag coefficient $C_{\mathrm{D}}$ is treated with the default value in ANSYS CFX of 0.44, $\rho_{\mathrm{m}} = \sum \alpha_k \rho_k$ is the density of the mixture and the interfacial area per unit volume $A_{\mathrm{m}}$ for the free surface model is calculated by $A_{\mathrm{m}} = |\nabla \alpha_k|$.

The time-dependent supercavitating turbulent flows are simulated using RANS equations together with the standard $k - \varepsilon$ turbulence model in the ANSYS CFX code. During the transient calculation, the convergence for each time step is achieved in 1000 iterations with the RMS (root mean square) residuals below $10^{-5}$. The high-resolution scheme is used for the advection terms, the central difference scheme is used for the diffusion terms and the second-order backward Euler scheme is used for the transient terms. The discrete continuity and momentum equations for the multi-phase flow field are solved together without iterations and corrections which improve the stability of the numerical scheme.

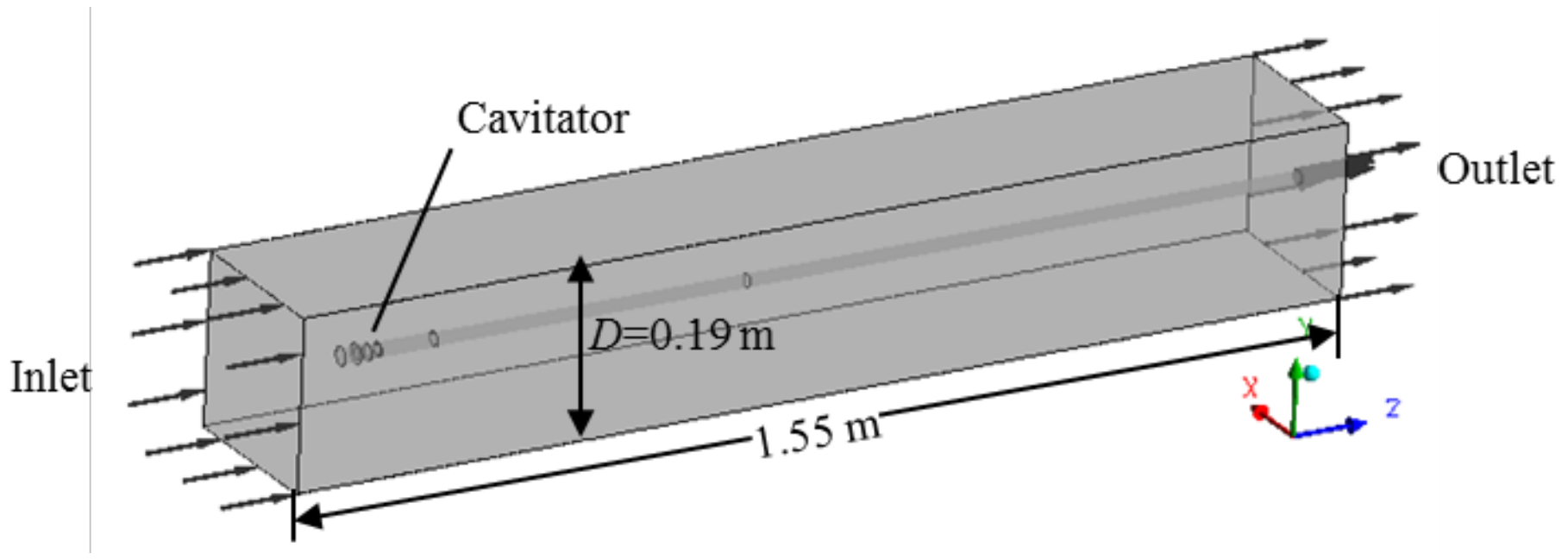

Fig.1: Three-dimensional computational domain ($d_C$ = 20 mm).

As shown in Fig. 1, a forward facing cavitator with the diameter of 20 mm, described by Kawakami and Arndt (2011), matching the model used in the experiments is employed to generate the cavity and the three-dimensional computational domain is 1.55 m in length with the same square cross section (0.19 m × 0.19 m) of the experiments. The streamwise and vertical direction corresponds to the *z*-axis and *y*-axis of the Cartesian coordinate system, respectively. The ventilation ports at the rear part of the cavitator are six circumferentially placed holes in the experiment. Such configuration is simplified to an annular band in simulation, which is an effective method to improve the local mesh quality and has been confirmed of its feasibility in supercavitation simulation in Cao et al. (2017). For the boundary conditions, the velocity components are used at the inlet that the streamwise velocity is $U_\infty = 8.5$ m/s with zero spanwise velocity (i.e., $x$-direction). The unsteadiness of the flow in the current paper is introduced through adding a vertical velocity term $v_g(t)$ at the inlet as a function of the pitching angle of the gust generator fitted to the experimental data (Kopriva 2008, Lee et al. 2013). No-slip conditions are employed on the other four boundaries of the computational domain except the inlet and outlet. A detailed description of the gust generator is provided in the next section. The pressure is adjusted at the outlet to match the cavitation number in experiments. The air flow rate ( $C_Q = 0.15$) is set at the ventilation port of the cavitator.

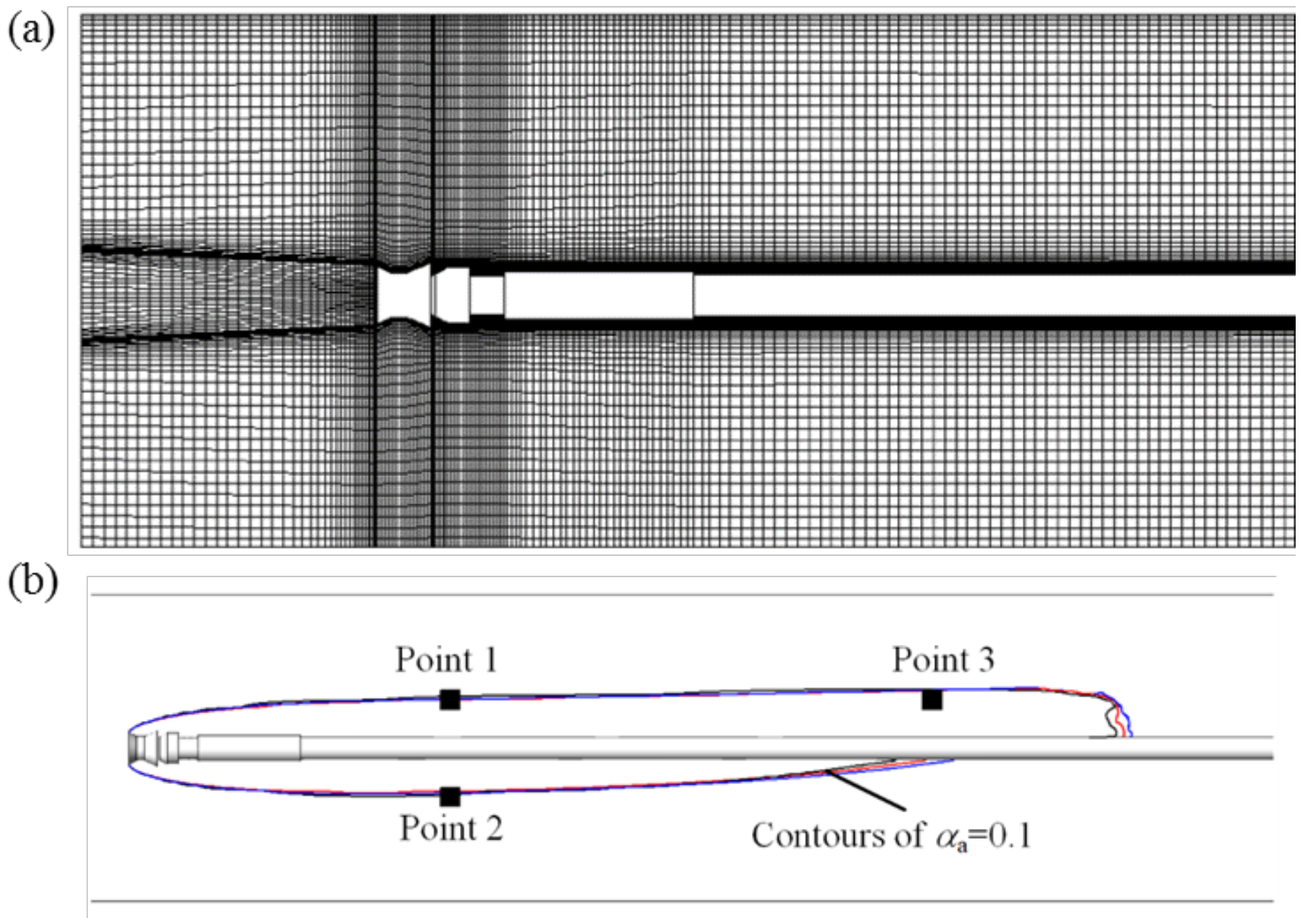

Fig. 2: (a) Structured mesh grids in the symmetry plane (i.e. *yz*-plane at *x*=0) and (b) supercavity shapes of three different grid resolutions corresponding to Table 1 where case 1 is in blue, case 2 is in red and case 3 is in black.

Table 1: Discretization uncertainty of time-averaged velocity ($u$) and absolute pressure ($p$) at point 1, 2 and 3 of Fig. 2(b). Point 1 and point 2 are at the maximum-diameter location and point 3 is near the cavity closure. The units for velocity and pressure are m/s and Pa, respectively.

| Mesh | Grid quantities | $u_1$ | $p_1$ | $u_2$ | $p_2$ | $u_3$ | $p_3$ |
|---|---|---|---|---|---|---|---|
| Case 1 | 2,644,817 | 9.04 | 58203.00 | 9.33 | 58305.60 | 8.61 | 58245.90 |
| Case 2 | 1,555,775 | 9.03 | 58484.60 | 9.25 | 58613.00 | 8.74 | 58537.30 |
| Case 3 | 915,161 | 9.01 | 58533.30 | 9.25 | 58674.10 | 8.66 | 58669.60 |

| | | | | | | |
|---|---|---|---|---|---|---|
| GCI | 1.29 % | 0.13 % | 0.02 % | 0.16 % | 3.67 % | 0.52 % |

Structured grids are generated in the computational domain as shown in Fig. 2(a), and grids around the cavitator are refined to allow a sufficient resolution to resolve the detailed physics of the supercavity. In order to determine an optimal grid resolution, three cases are tested under the flow conditions of $Fr = 18.7$, $C_{\mathrm{Q}} = 0.15$, $d_c = 20\ \mathrm{mm}$ and $\sigma_{\mathrm{steady}} = 0.20$. Three grid quantities are determined with a constant grid refinement ratio $r = 1.3$ in all three directions according to Table 1. As shown in Fig. 2(b), cavity maximum diameter locations (point 1 and 2) and a point near cavity closure region (point 3) are chosen as the monitoring points for the comparison of results from different cases. The grid convergence index (GCI) of velocity and static pressure at monitoring points (Roache 1993) is introduced to estimate the uncertainty. As shown in Table 1, the uncertainty estimated by GCI method along the streamwise direction with has a value less than 5 % which demonstrates that the simulation results are almost independent of the grid resolution. Additionally, the comparison results of cavity shape (defined the contours of air volume fraction equals to 0.1) under three grid quantities are closely matched with each other with small differences at the cavity closure region (Fig. 2b). Note that the further refinement of grids will lead to substantial increasing of computational cost and the instability of the numerical simulation. Therefore, we choose the case 2 as the final mesh in the simulation corresponding to about 1.6 million elements.

### 2.2 Experimental setup

The validation experiments are conducted in a high-speed water tunnel in the Saint Anthony Falls Laboratory at University of Minnesota with details provided in the past studies (Lee et al. 2013, Karn et al. 2015, Karn et al, 2016, Shao et al. 2018 and Wu et al. 2019). It is a closed recirculating facility and the length of the horizontal test-section is 1.20 m long and the same cross section with the simulation.

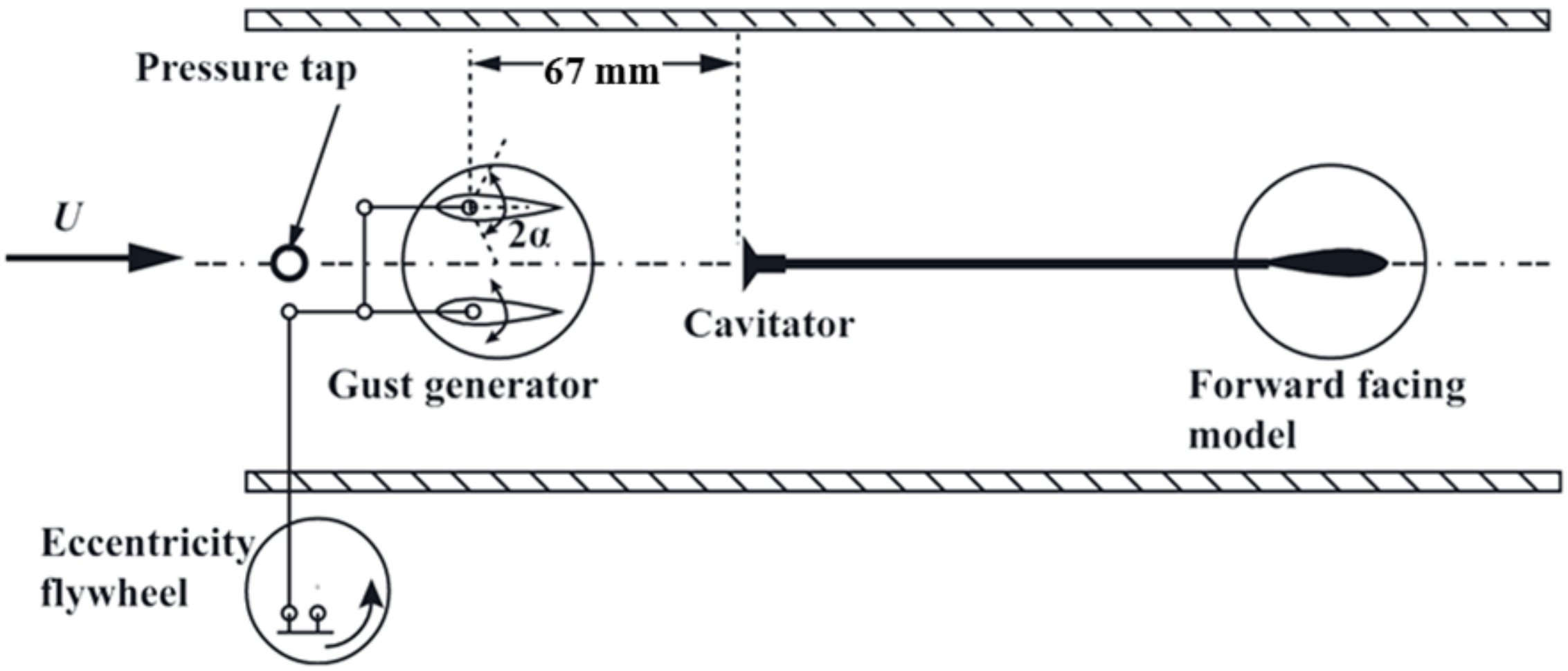

Fig. 3: Schematic of the experimental setup including the gust generator and the forward facing cavitator (adapted from [12]).

As shown in Fig.3 of the experimental setup, two oscillating NACA0020 hydrofoils with chord length of 40 mm are placed 67 mm upstream of the cavitator to generate a periodical varying unsteady flow. The hydrofoils can oscillate in phase driven by a pivot arm which is connected to an eccentricity flywheel outside the tunnel. It is worth noting that unlike the simulation, the flapping motion of the hydrofoils provides a test section pressure variation associated with varying blockage ratios of the hydrofoils besides the vertical velocity term. Instantaneous velocities of the periodic gust flows are measured by Laser Doppler Velocimetry in (Korpriva et al. 2008) with the vertical velocity component fitted to a sinusoidal function, i.e. $v_{\mathrm{g}}(t) = v_{\mathrm{gmax}} \sin(2\pi f_{\mathrm{g}} t)$, where $v_{\mathrm{gmax}}$ is the maximum vertical velocity (i.e., amplitude) measured near the leading edge of the stationary hydrofoils controlled by the angle of attack ($\gamma_{\mathrm{g}}$) of two hydrofoils, and $f_{\mathrm{g}}$ is the frequency of periodic flows (i.e., wavelength) corresponding to the rotational speed of the motor. A forward facing model that matches that of the simulation is employed in experiments to generate a ventilated supercavity with smooth surface. The air flow is injected from the ventilation ports at the cavitator and controlled by a mass flow controller. During the experiments, two sensors are employed to measure the pressure before the gust generator and within the cavity 64 mm behind the cavitator. The pressure sensors are sampled at 1000 Hz with the uncertainty around ±100 Pa. The high-speed imaging is synchronized with the pressure measurements through the LabVIEW software. Detailed experimental setup and data acquisition procedures can be found in (Shao et al. 2018).

### 2.3 Validation of the numerical approach

First, the ventilated supercavity in the steady incoming flow is used to validate the present numerical approach. The conditions are $Fr = 18.7$, $C_{\mathrm{Q}} = 0.15$, $d_c = 20$ mm and $\sigma_{\mathrm{steady}} = 0.20$ measured in experiments, under which the cavity has a twin-vortex closure type. The numerical simulation is carried out with the same boundary conditions as the experiment by adjusting the outlet pressure to match the cavitation number. The experimental and simulated supercavity geometries are compared in Fig. 4 with two geometrical parameters proposed by Brennen (1969), i.e. maximum diameter ($D_{\mathrm{max}}$) and half-length ($L_{1/2}$) of the supercavity which is the horizontal distance between the cavitator and $D_{\mathrm{max}}$. It is shown that the simulated cavity has maximum diameter of 61 mm and half-length of 200 mm, while the experimental results are 62 mm and 204 mm respectively. The predicted supercavity shifts upward at the rear part due to the buoyancy effect and the twin-vortex closure is successfully predicted as shown in Fig. 4(b), which is also observed in experiments.

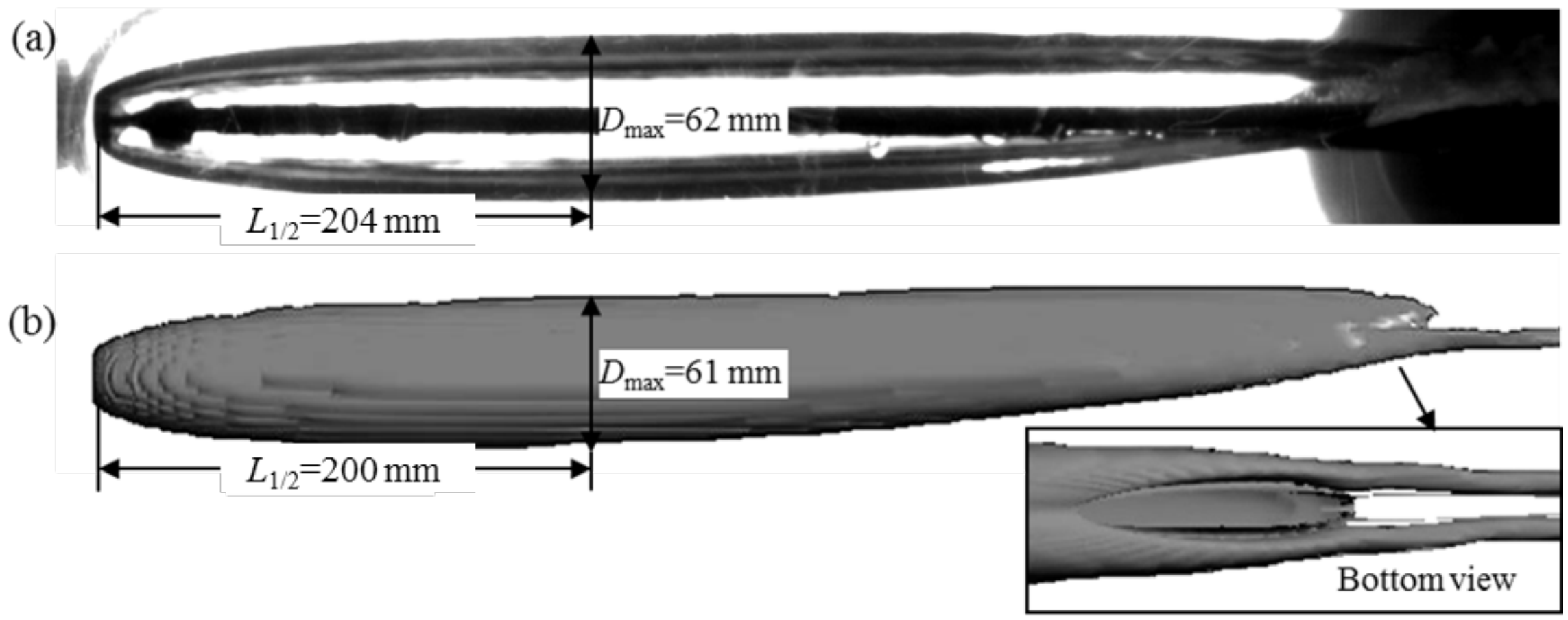

Fig. 4: Comparison of (a) the experimental supercavity and the simulated supercavity with the bottom view showing the twin-vortex closure under steady condition of $Fr = 18.7,\ C_Q = 0.15, d_c = 20$ mm and $\sigma_{steady} = 0.20$

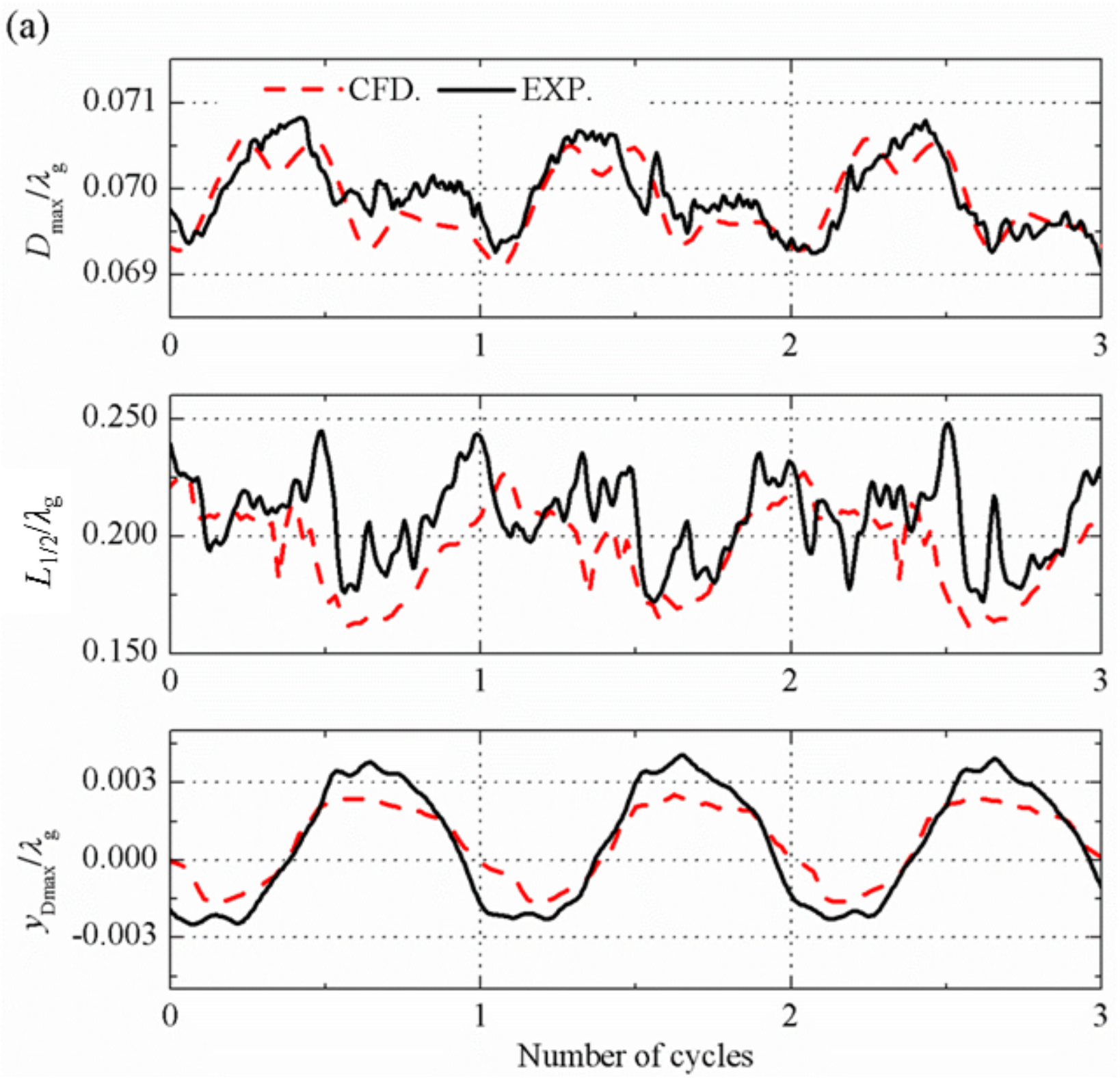

Fig. 5: Comparison of (a) the cavity geometry and (b) the instantaneous pressure variation under unsteady conditions between the experiment and the simulation. The flow condition is $Fr = 18.7,\ C_Q = 0.15, d_c = 20$ mm and $\sigma_{steady} = 0.20$ with $\gamma_g = 4°$ and $f_g = 10$ Hz. Note that for $\sigma$, the simulation result uses the *y*-scale labeled on the left side and experiment result uses the right side one.

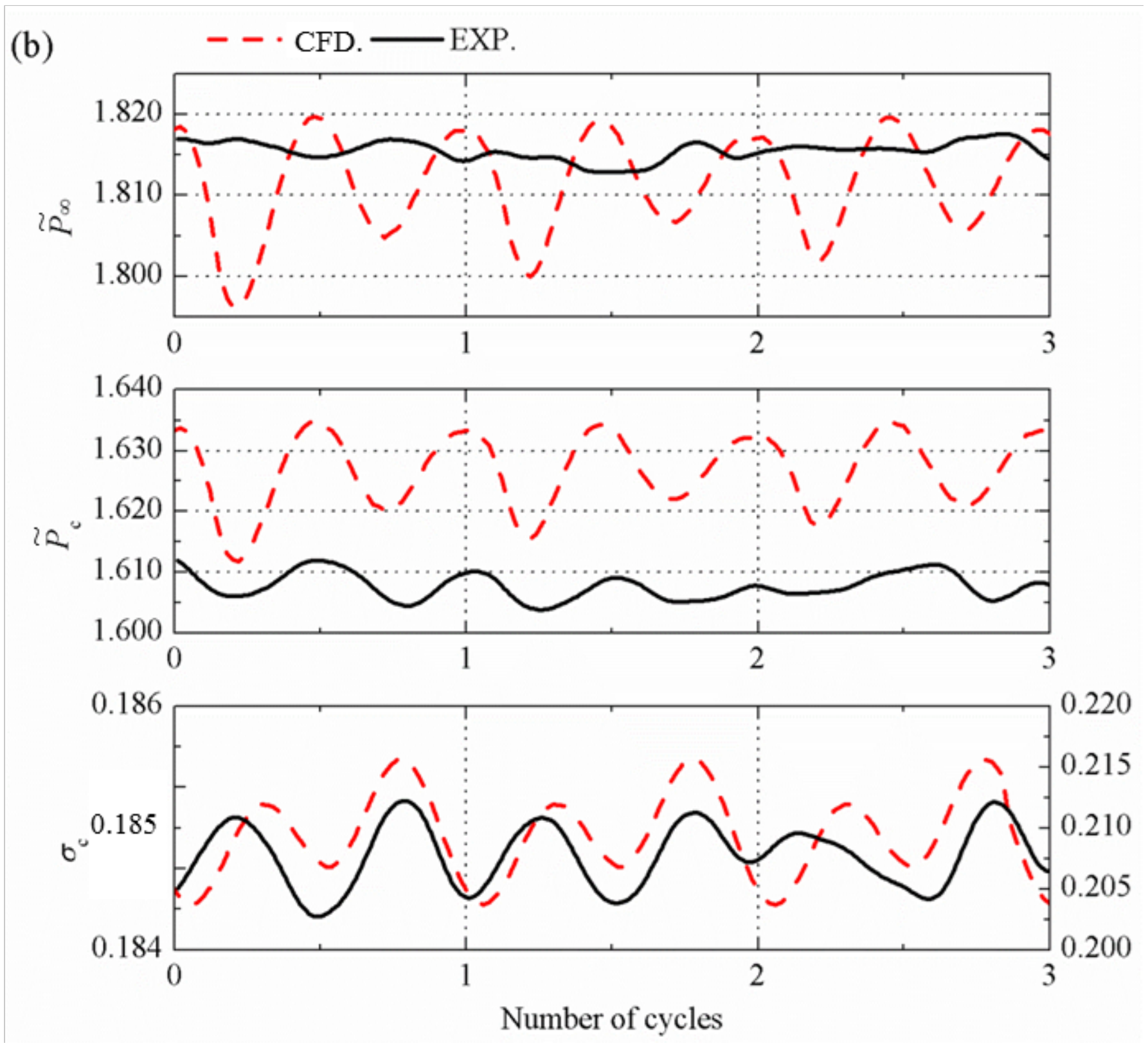

Fig. 5 (continued)

Second, the numerical approach is validated by the ventilated supercavity in the unsteady flow settings. The same conditions with the steady validation adding a periodical flow set as $f_g = 10$ Hz and $\gamma_g = 4°$ are employed in both the numerical simulation and the experiment. The numerical results (CFD, for short) and experimental data (EXP, for short) are shown in Fig. 5(a) by comparing the $D_{\max}$ and its locations ($L_{1/2}$,$y_{D_{\max}}$). The location of maximum diameter is defined in the standard *yz*-coordinate system with the origin at the cavitator center. All parameters are normalized by the gust wavelength $\lambda_g = U_\infty / f_g$. The $L_{1/2}$ and $y_{D_{\max}}$ in the experiments are measured through an automatic image processing technique described by Lee et al. (2013) with an uncertainty around 8 %. It is depicted that $D_{\max}$ and its location are fluctuated periodically both in simulation and experiments. The simulated $D_{\max}$ and $y_{D_{\max}}$ vary synchronously with the experiments. In the simulation, the $D_{\max}$ fluctuates only about 1% around its mean, while the $L_{1/2}$ fluctuation is about 16.5 %, matching closely with the experiment values (i.e., 1.1 % for $D_{\max}$, and 17.0 % for $L_{1/2}$, respectively). However, the fluctuation of $y_{D_{\max}}$ from the experiment is about 30% higher than that from the simulation. We attribute such discrepancy to the differences in the unsteady flow setup between the simulation and experiment. Specifically, the varying blockage associated with the hydrofoil pitching in the experiments leads to the stronger shape fluctuation of the cavity.

The supercavity pressure fluctuation in a gust cycle is also compared across the experiment and the simulation (Fig. 5b). The test section and cavity pressures captured in the simulation and the experiment are normalized by the dynamic pressure of the water flow. The experiment and simulation show similar periodical variation of cavity pressure and cavitation number. Additionally, both the experiment and the simulation show double peaks of the test section and cavity pressure variations within a gust cycle, corresponding to two instances of the hydrofoils with the maximum flapping angles. It is worth noting that the amplitude of the fluctuation in the normalized test section pressure is around 0.2% which is larger than the uncertainty of the pressure measurement around 0.1% corresponding to 100 Pa. In the simulation, the $p_{\infty}$ fluctuates about 0.5 % around its average, and $p_{\mathrm{c}}$ has a fluctuation around 0.6 %, in comparison to their counterparts of the experiments results both around 0.2 %. Nevertheless, the fluctuation of $\sigma_{\mathrm{c}}$ for the simulation is around 0.3 % around its average value and showing a slight phase difference with the experimental result fluctuating at the amplitude of 2.4 %. The varying minor loss of the tunnel corresponding to different phases of the flapping foils attenuates the amplitudes of pressure fluctuations during the experiments. This minor loss also leads to the varying tunnel velocity and a local minimum of $p_{\infty}$ at the minimum angle of attack of the foils and a local maximum corresponding to the maximum angle of attack of the foils (note that the tunnel is vented to the atmosphere during the experiments). Therefore, an out-of-phased response of the $p_\infty$ and $p_c$ to the phase change of flapping foils occurs and eventually causes the enhanced $\sigma_c$ fluctuations during the experiments.

## 3. Results

Based on the validations in section 2.3, different unsteady flow conditions listed in Table 2 are simulated with periodic gust flows corresponding to different gust generator setups ($f_g$ and $\gamma_g$) at fixed $d_c = 20$ mm, $Fr = 18.7$ , $C_Q = 0.15$, and $d_c = 20$ mm .The transient supercavity behavior and the cavity shape deformation are provided during one cycle for one the case of AOA8f10 state ($f_g = 10$ Hz and $\gamma_g = 8°$), and the effects of $\gamma_g$and $f_g$ are discussed subsequently. For all the simulations, the cavity boundaries are defined as the iso-surface of the air volume fraction of 10%

Table 2: Conditions of the simulated states with the vertical velocity component calculated by $v_g(t) = v_{gmax}\sin(2\pi f_g t)$.

| **state** | **angle of attack (°)** | $\boldsymbol{f_g}$ **(Hz)** | $\boldsymbol{v_{gmax}}$ **(m/s)** |
|---|---|---|---|
| AOA8f10 | 8 | 10 | 0.227 |
| AOA6f10 | 6 | 10 | 0.327 |
| AOA4f10 | 4 | 10 | 0.448 |
| AOA8f5 | 8 | 5 | 0.465 |
| AOA8f1 | 8 | 1 | 0.5 |

### 3.1 Transient supercavity behaviors

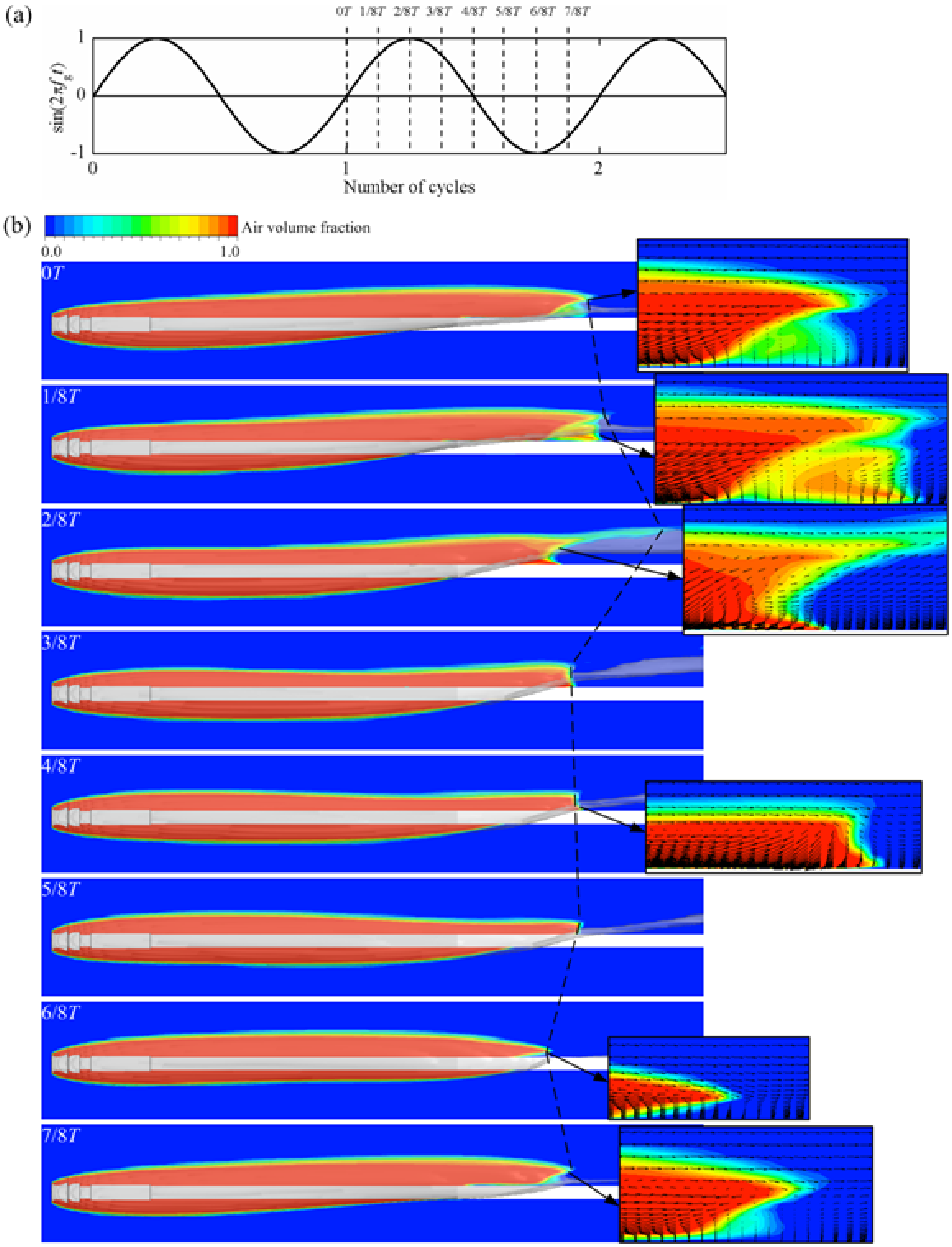

Fig. 6: The variation of supercavity during a gust cycle at AOA8f10 state ($\gamma_g = 8°$ and $f_g = 10$ Hz). The contour shows the air volume fraction ($\alpha_a$) at the symmetry plane, and the transparent iso-surface corresponds to $\alpha_a = 0.1$. Inset figures show the contour of air volume fraction overlaid with the velocity vector field near the closure region.

Fig. 6 shows the supercavity variation at eight instants during a gust cycle for AOA8f10 state ($f_{\mathrm{g}} = 10$ Hz and $\gamma_{\mathrm{g}} = 8°$). Inset figures of $\alpha_{\mathrm{a}}$ overlaid with the velocity vector field near the closure region are also provided at selected instants. At $t = 0\,T$, where $T$ is the period of the foil flapping, the supercavity encompasses the cavitator, grows along the strut and closes with two hollow vortex tubes. The cavity body shows an upward shift at the rear region due to the buoyancy effect. In accordance with periodic flow unsteadiness, the supercavity body fluctuates periodically in the vertical direction at the symmetry plane, especially at its closure region. In addition, it exhibits a periodical growth/shrink in the streamwise direction with the cavity length illustrated by the dash lines. Specifically, during $t = 0{\sim}2/8\,T$, the cavity body extends downstream, and meanwhile the two hollow vortex tubes moves upward and enlarges. Both the cavity length and the diameter of vortex tubes reach the maximum values at $t = 2/8\,T$. Subsequently, the cavity body begins to shrink along the ventilation pipe until the minimum length at $t = 6/8\,T$. As a result, the twin vortex tubes reduce in diameter and then disappear at $t = 6/8\,T$, transitioning to a re-entrant jet closure. Finally, both the cavity and twin vortex tubes grow again at $t = 7/8\,T$ followed by a new cycle. Moreover, a low-$\alpha_{\mathrm{a}}$ portion is clearly observed near the cavity closure at $t = 0\,T$ and $t = 1/8\,T$, indicating an entrant-jet type of flow. Such flow declines during $= 2/8{\sim}6/8\,T$ , and reoccurs at $t = 7/8\,T$. As shown in the close-up view in Fig. 6 at $t = 2/8\,T$, near the cavity closure region, the majority of the flow field within the vertical extent of the cavity is occupied by a reverse flow except in the internal boundary layer formed at the water-gas interface and the flow in vortex tubes. Wu et al. (2019) observed such internal boundary layers in the upstream portion of a supercavity generated by a backward-facing cavitator (i.e. the cavitator with the mounting strut in front of the cavitator) and postulated the similar internal flow field near the closure region. Additionally, in comparison to Wu et al. (2019), the simulation uses different cavitator configuration with a ventilation pipe behind the cavitator. As a result, as depicted in the close-up views at $t = 2/8\,T$ and $t = 4/8\,T$, a velocity boundary layer develops on the surface of the ventilation pipe with strong reverse flow.

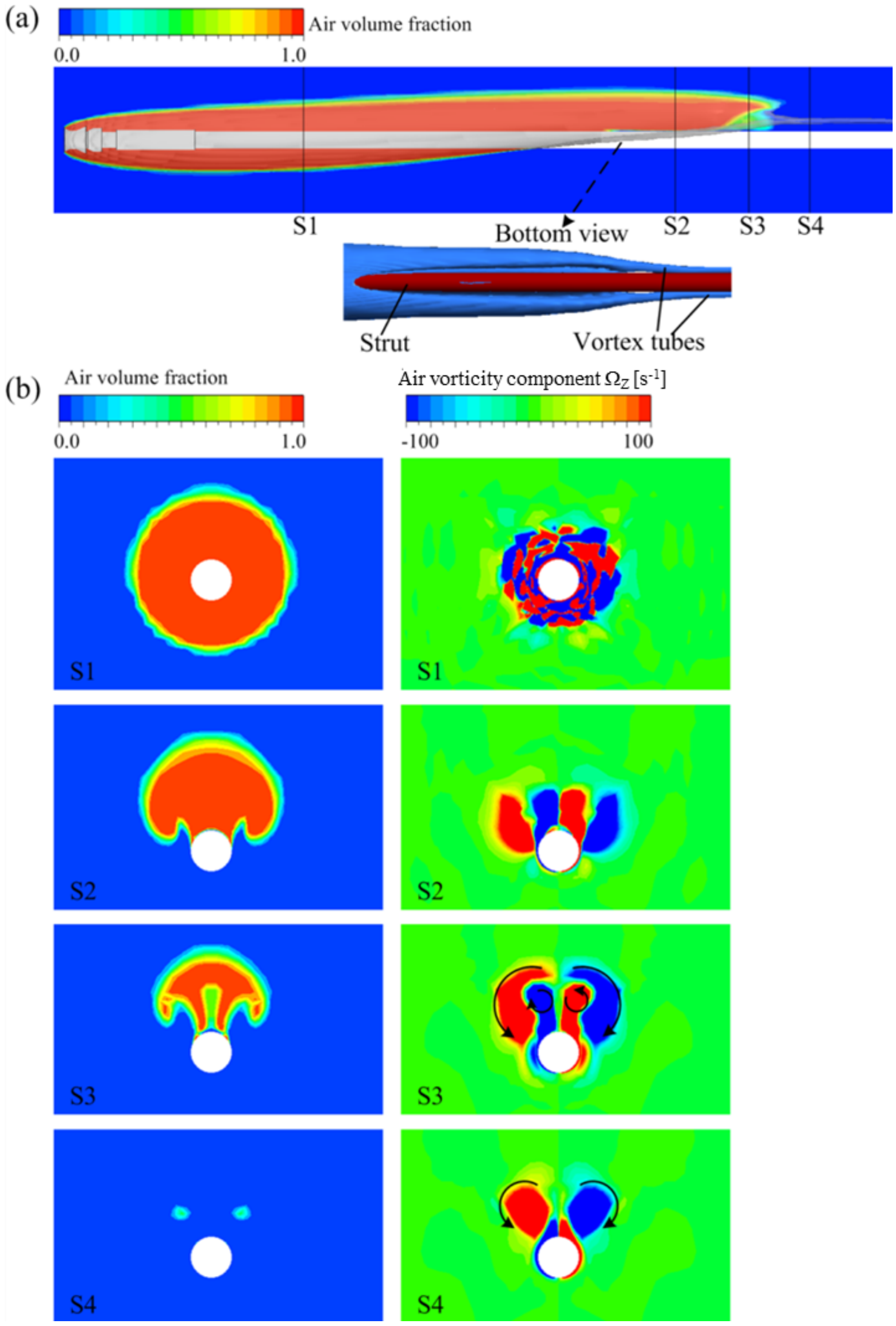

Fig. 7: Distributions of air volume fraction and streamwise vorticity component $\Omega_z$ at four cross sections (S1, S2, S3, S4) at $t = 0T$ for the AOA8f10 state ($\gamma_g = 8°$ and $f_g = 10$ Hz).

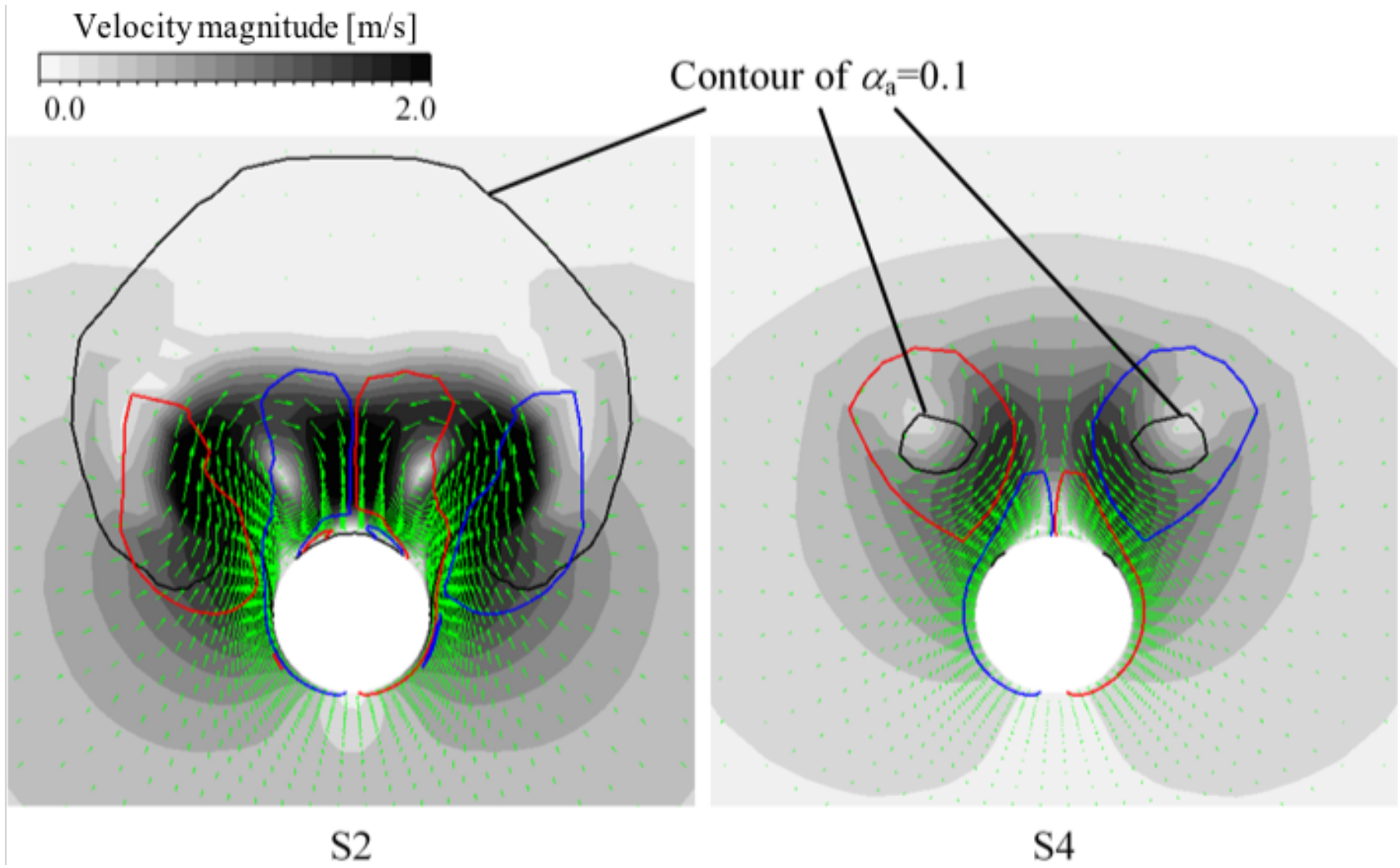

Fig. 8: Air velocity on S2 and S4 cross sections of the cavity at $t = 0T$ for the AOA8f10 state ($\gamma_g = 8°$ and $f_g =$ 10 Hz) overlaid with air fraction contour corresponding to air volume fraction of 10% (marked in black) and vorticity contours (clockwise rotation in blue and anti-clockwise rotation in red).

The variation of internal flow field along the supercavity is further examined with distribution of air volume fraction and vorticity at cross-section plane at different streamwise locations as shown in Fig. 7. The distributions of $\alpha_a$ and the streamwise vorticity component ($\Omega_z = \partial u_y/\partial x - \partial u_x/\partial y$) are illustrated in four cross sections (termed as S1, S2, S3 and S4). Note that the S1 is the location of the cavity maximum diameter and others are near the closure region. The supercavity is symmetric at the S1 section and so is the internal flow depicted by the vorticity distribution. At S2 and S3 sections, the buoyancy effect causes the cavity to curve upwards and its cross section develops to a "crescent" shape with high volume fraction as reported in (Wang et al. 2015), which generates two pairs of counter-rotating vortices at S2 and S3 sections. At the section S4, the air volume fraction decreases dramatically, and the outer pair of vortices becomes dominant with the inner pair of vortices located around the ventilation pipe. Such vortex pairs are associated with the velocity field at the cross section as depicted in Fig. 8, which shows the air contours corresponding to the air volume fraction about 10 % (marked in black) along with the vortex outlines (marked in blue and red). The outer vortex pair is generated from the strong velocity gradient near the water-gas interface of the cavity, while the inner pair of vortices is a result of the velocity boundary layer developed along the mounting strut, which is not present in the internal flow experiment from Wu et al. (2019) due to different cavitator configurations.

### 3.2 Cavity shape deformation

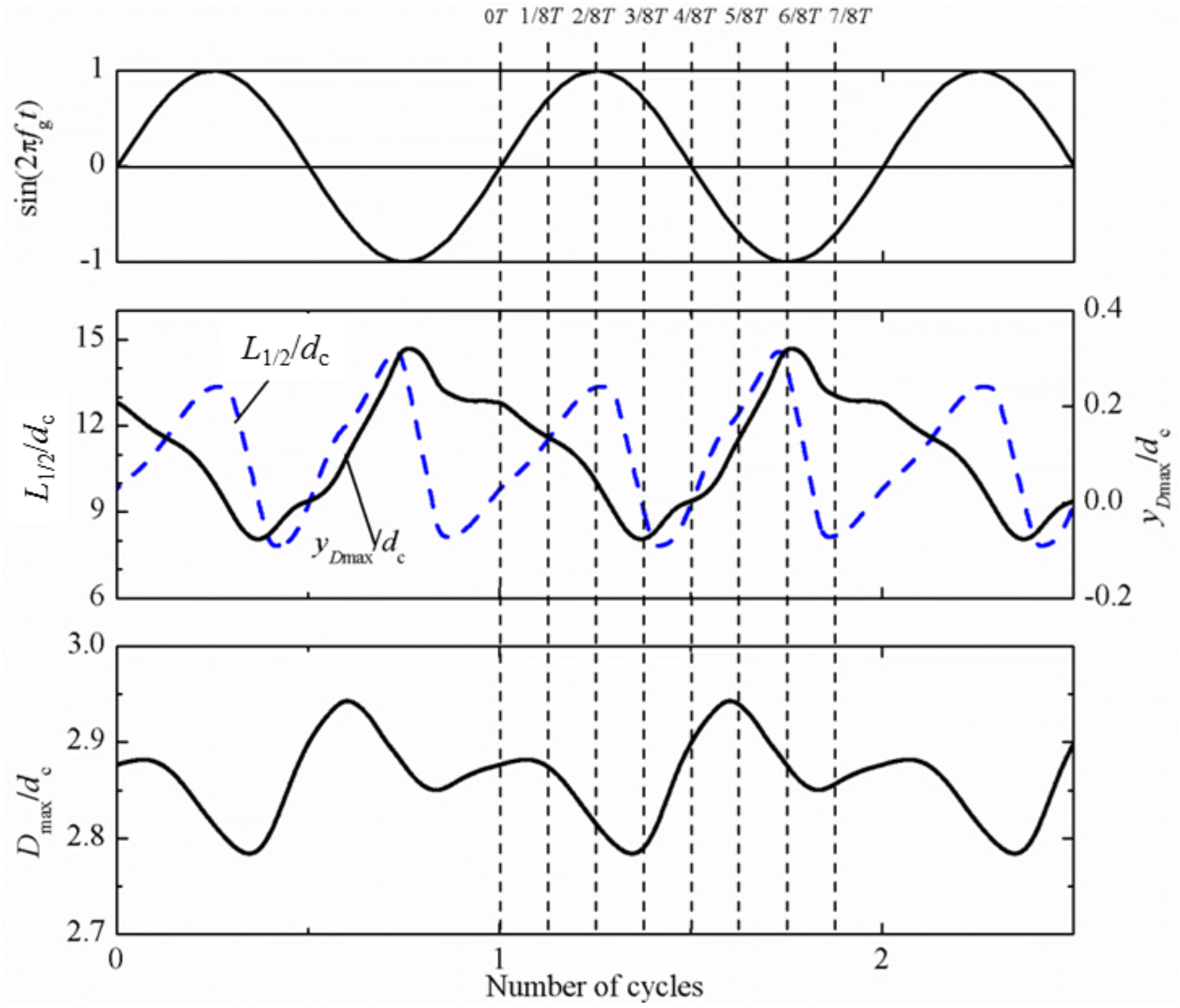

Fig. 9: The periodic variations of the supercavity maximum diameter ($D_{\max}$) and its locations ($L_{1/2}$, $y_{D_{\max}}$) at AOA8f10 state ($\gamma_g = 8°$ and $f_g = 10$ Hz).

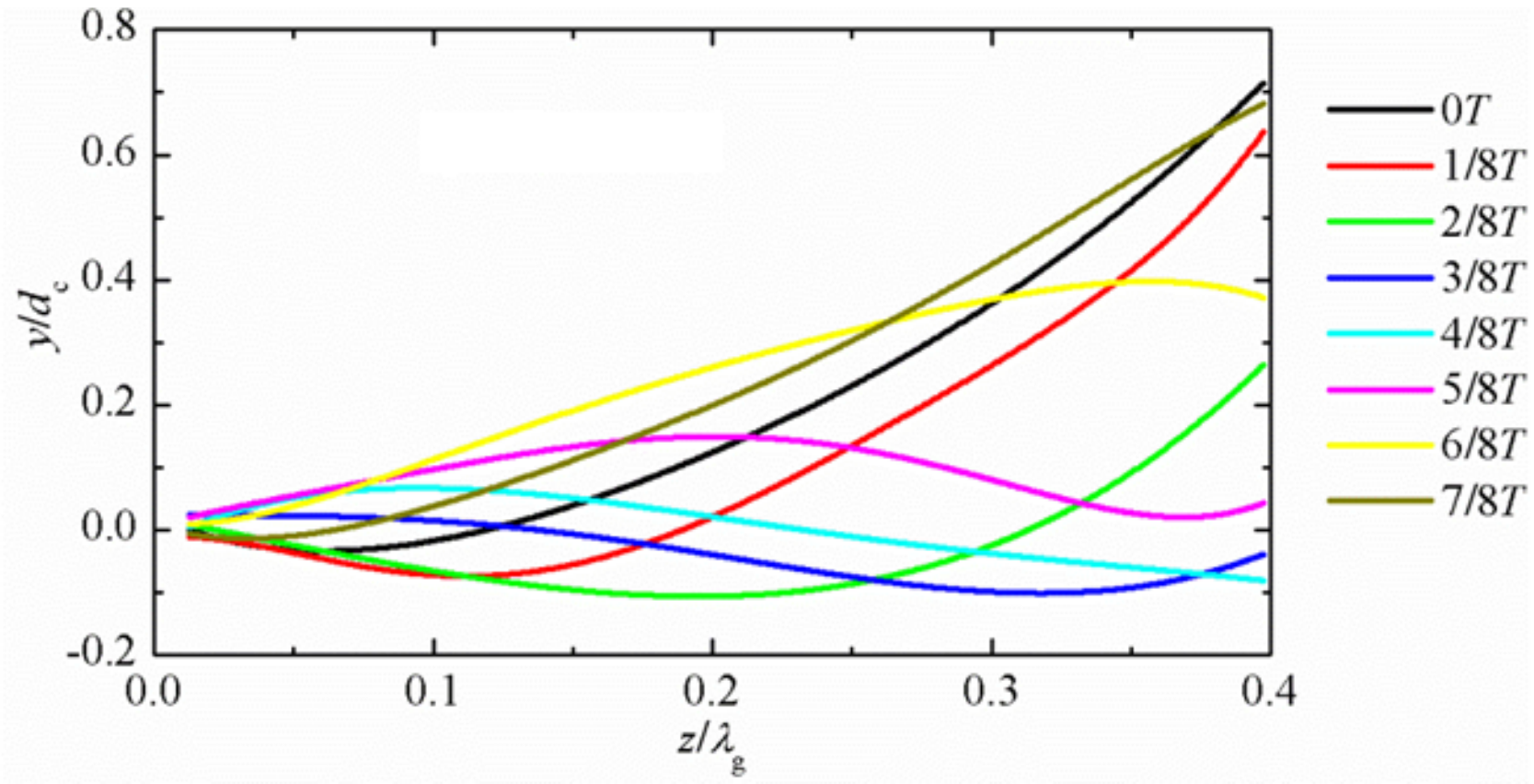

Fig. 10: Variations of the supercavity centerline during a gust cycle at AOA8f10 state ($\gamma_g = 8°$ and $f_g = 10$ Hz). Each color in the legend corresponds to one of the eight instants shown in Fig. 9.

Firstly, the variation of the cavity overall geometry is depicted as Fig. 9 by showing the time-dependent $D_{\max}$ of the supercavity and its locations ($L_{1/2}, y_{D_{\max}}$). The cavity half length $L_{1/2}$ varies at twice of the $f_g$ with the peaks located at the extrema of the maximum hydrofoil vertical speeds, while the frequency of $D_{\max}$ and $y_{D_{\max}}$ stays the same as that of the gust flow. Additionally, the $L_{1/2}$ oscillates at an amplitude around 33.0 % comparing to its average while $D_{\max}$ is around 3.5 %. Comparatively, the $y_{D_{\max}}$ has a fluctuating amplitude around 64.3 % relative to its maximum value. The variation of cavity morphology is characterized by the change of its instantaneous centerline at the symmetric *yz*-plane extracted from its iso-surface at eight consecutive time instances ($t = 0 \sim 7/8\,T$) in a gust cycle. It is worth noting that the cavity is located in the wavy state described in the previous experimental investigation (Shao et al. 2018). The streamwise position *z* is normalized by the gust wavelength (i.e. $\lambda_g = U_\infty/f_g = 0.85$ m) and vertical location is normalized by the cavitator size ( $d_c$) in Fig. 10. The cavity centerline exhibits wavelike deformation in the unsteady flow similar to the experimental results. It is worth noting that due to the buoyancy, the cavity has relatively stronger fluctuation in its centerline after $L_{1/2}$. The minimum vertical position of the centerline is advected at the speed of 96.5 % of the free stream flow speed. Additionally, the amplitude of fluctuation of the cavity centerline in *y*-direction comparing to its neutral location is around 6 mm or 0.7 % relative to the wavelength which is in the same order of the wave amplitude (0.8 % of relative to the wavelength). Similar cavity geometrical results (i.e., cavity deformation advection speed and degree of surface deformation) are observed from the experiments under same flow conditions (Shao et al. 2018). Therefore, the numerical settings presented in this paper is able to capture the geometrical behaviors of wavy state cavity in the unsteady flow.

### 3.3 Pressure distributions and ventilated cavitation number

The instantaneous cavitation number is calculated from the difference between test section pressure and cavity pressure normalized by the dynamic pressure of the incoming flow as shown in Fig. 11. Compared to their averaged values, dimensionless test section pressure ($P_\infty$) has a fluctuation of 1.9 % around its mean, and $P_c$ fluctuates slightly stronger at 2.2 %. The cavity pressure has a phase delay of 0.014 s with respect to the phase of hydrofoils is associated with the propagation of the perturbation induced by the flapping hydrofoils.

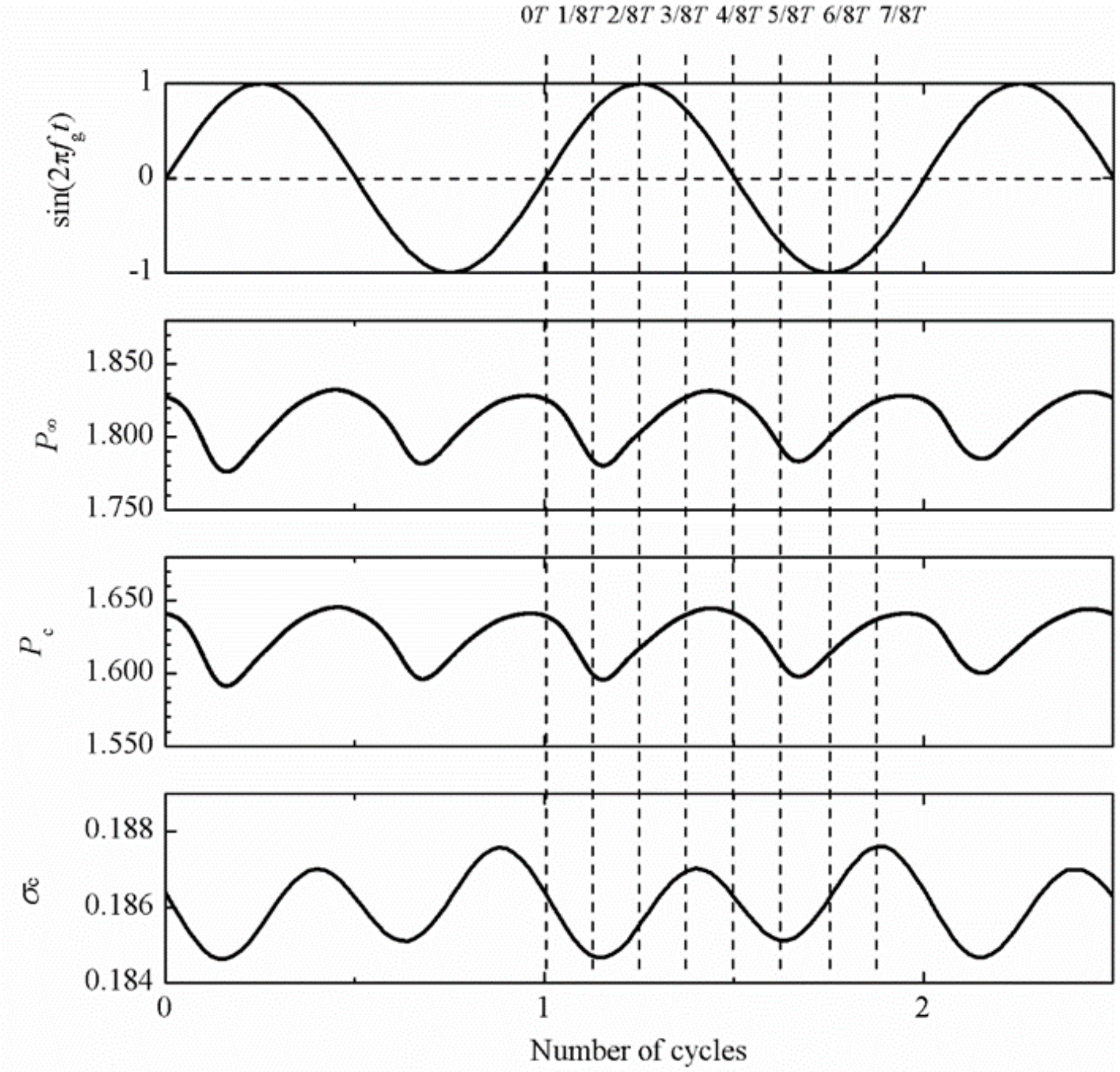

Fig. 11: The instantaneous variation of non-dimensionalized test section pressure cavity pressure and cavitation number at AOA8f10 state ($\gamma_g = 8°$ and $f_g = 10$ Hz).

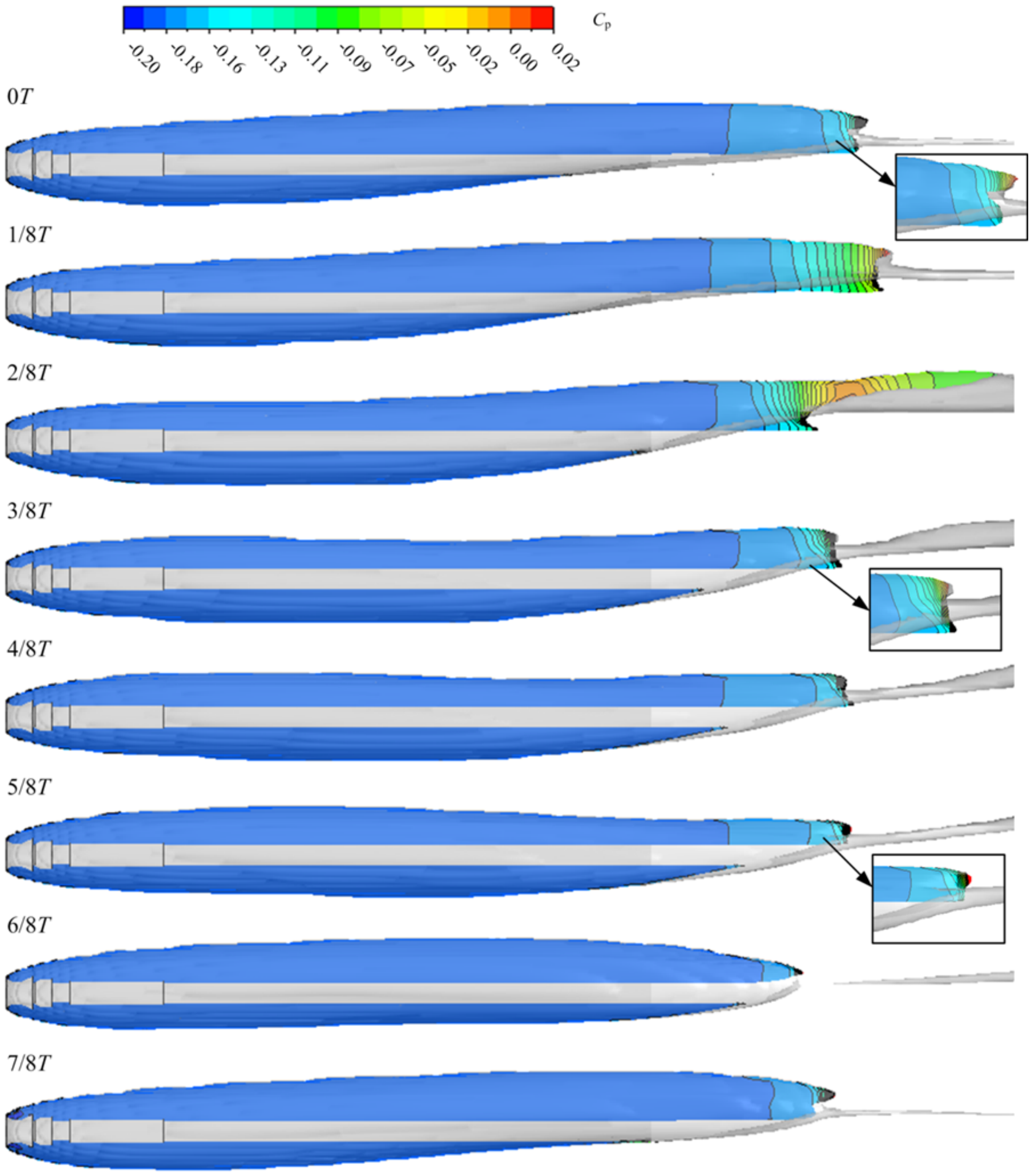

Fig. 12: The distribution of pressure coefficient inside the cavity during a gust cycle at AOA8f10 state ($\gamma_g = 8°$ and $f_g = 10$ Hz). The contour shows the air volume fraction ($\alpha_a$) at the symmetry plane, and the transparent iso-surface corresponds to $\alpha_a = 0.1$.

Fig.12 shows the pressure coefficient, $C_p = 2(p - p_\infty)/(\rho_w U_\infty^2)$, calculated at eight instants during a gust cycle at AOA8f10 state, to give an insight into the pressure distribution inside the cavity. Similar to the steady case (Cao et al. 2017), in a gust cycle, the pressure distribution remains largely uniform in the majority portion of the cavity at the front with a variation of pressure gradient near the cavity closure. Near the closure region, the pressure shows a steeper gradient along the streamwise direction at $t = 1/8\,T$ and $2/8\,T$ compared to other phases in a cycle, as the low-pressure air inside the cavity rapidly increases its pressure to match the high pressure of the water after being discharged from the hollow vortex tube. Such change of pressure gradient at the closure leads to the closure modes variation as reported in Karn et al. (2015) and Shao et al. (2018). Specifically, as the closure mode transitioning from reentrant jet closure (corresponding to $t = 6/8\,T$) to twin vortex closure ($t = 2/8\,T$), the pressure at the closure of the cavity first decreases sharply and then increases as the air moves downstream into the vortex tube.

Table 3: The pressure variations ($\Delta p$) inside and outside the closure ($\tilde{P}_{\text{in}}, \tilde{P}_{\text{out}}$) at eight instants during one cycle illustrated in Fig. 11.

| **Time** | $\tilde{\boldsymbol{P}}_{\mathbf{in}}$ | $\tilde{\boldsymbol{P}}_{\mathbf{out}}$ | $\Delta\tilde{\boldsymbol{P}}$ |
|---|---|---|---|
| 0/8$T$ | 0.20 | 1.87 | 0.20 |
| 1/8$T$ | 0.15 | 1.86 | 0.13 |
| 2/8$T$ | 0.09 | 1.85 | 0.12 |
| 3/8$T$ | 0.17 | 1.90 | 0.17 |
| 4/8$T$ | 0.19 | 1.91 | 0.19 |
| 5/8$T$ | 0.28 | 1.96 | 0.28 |
| 6/8$T$ | 0.36 | 2.01 | 0.36 |
| 7/8$T$ | 0.21 | 1.88 | 0.21 |

The connection between the pressure variation and the change of closure modes during the gust cycle can be explained using the framework proposed in Karn et al. (2016) by introducing a differential pressure term, $\Delta\tilde{P} = (p_{\text{out}} - p_{\text{in}})/(1/2\,\rho_{\text{w}} U_\infty^2) = \tilde{P}_{\text{out}} - \tilde{P}_{\text{in}}$, where $\tilde{P}_{\text{in}}$ and $\tilde{P}_{\text{out}}$ represent the dimensionless static pressure inside and just outside the cavity at the closure, respectively. The $\tilde{P}_{\text{in}}$ is calculated to be the average along a vertical line corresponding to a mean air volume fraction of 0.9 in the vicinity of cavity closure location (marked by the black dashed line in Fig. 6), whereas $\tilde{P}_{\text{out}}$ is the averaged pressure in the supercavity wake. As shown in Table 2, the $\Delta\tilde{P}$ is 0.36 for a re-entrant jet type closure at $t = 6/8\,T$ which is considerably larger than the corresponding value of 0.12 for a twin vortex closure at $t = 2/8\,T$, providing further support to the pressure criterion proposed by Karn et al. (2016).

### 3.4 Effect of the angle of attack

The effect of the angle of attack on cavity behavior is investigated through the simulation under three angles of attack, i.e., $\gamma_{\text{g}} = 4°, 6°, 8°$, with a fixed gust frequency at $f_{\text{g}} = 10$ Hz at the same inlet flow speed and ventilation rate. Fig. 13 shows the deformation of the cavity centerline in a gust cycle upon different angles of attack. Compared the results shown in Fig.10 (the result is also added in the Fig. 13 for better comparison), under lower angles of attack, the cavity still exhibits wavelike undulation. However, with the reduction of $\gamma_{\text{g}}$, there is a corresponding decrease of the undulation amplitude along the entire cavity. At $\gamma_{\text{g}} = 8°$ , the cavity center line has the maximum undulation at $t = 0\,T$ at the end of the cavity (i.e. corresponding to $z/\lambda_{\text{g}} = 0.4$ in the figure). Remarkably, as the $\gamma_{\text{g}}$ changes from 8° to 4°, the undulation amplitude of the cavity centerline at $t = 6/8\,T$ gradually catches up and then exceeds that at $t = 0\,T$ at the end of the cavity.

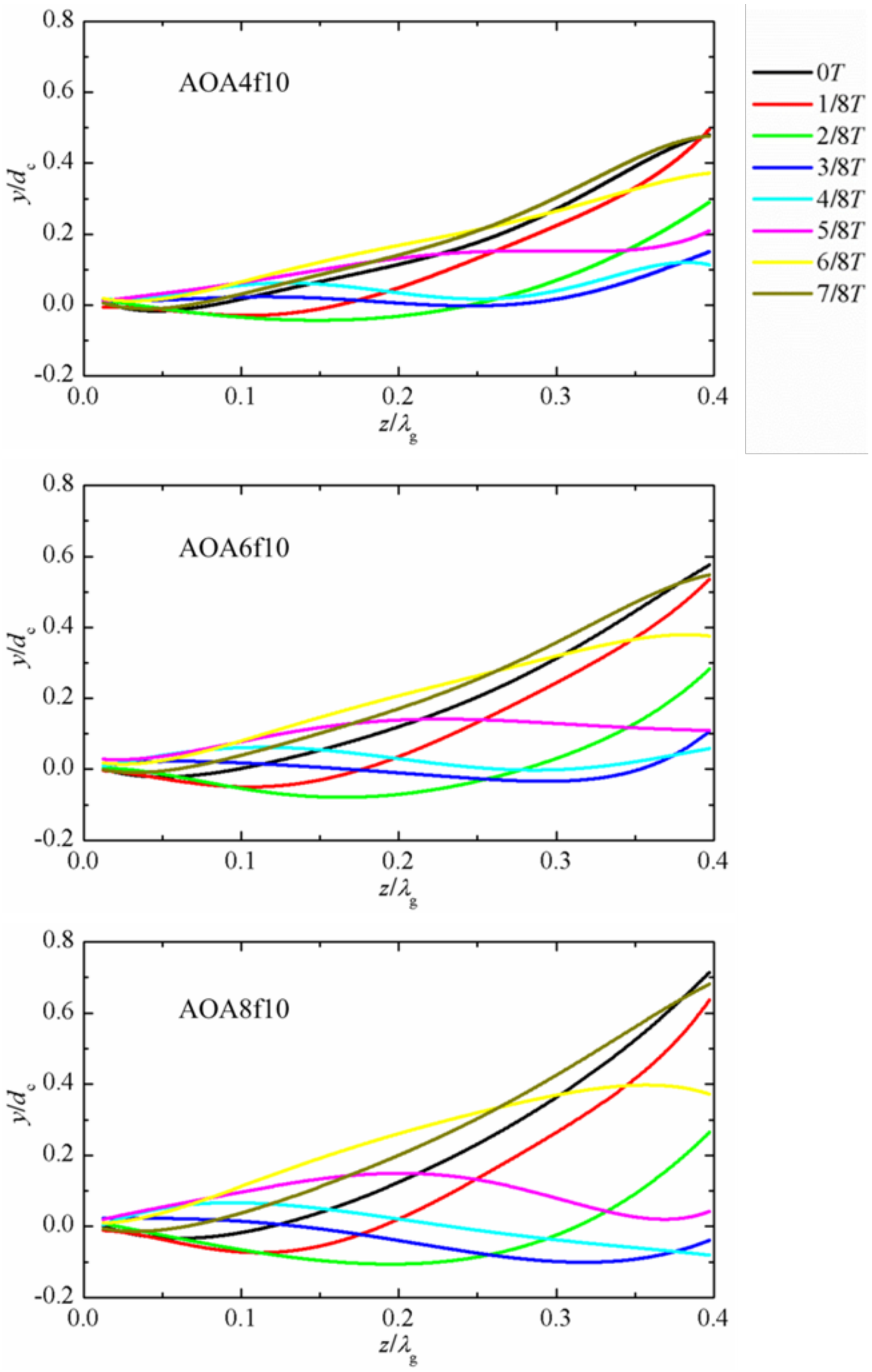

Fig. 13: Effects of the angle of attack ($\gamma_{\mathrm{g}}$) on the variations of supercavity centerlines under the pitching frequency $f_{\mathrm{g}} = 10$ Hz and the angles of attack $\gamma_g = 4°, 6°, 8°$.

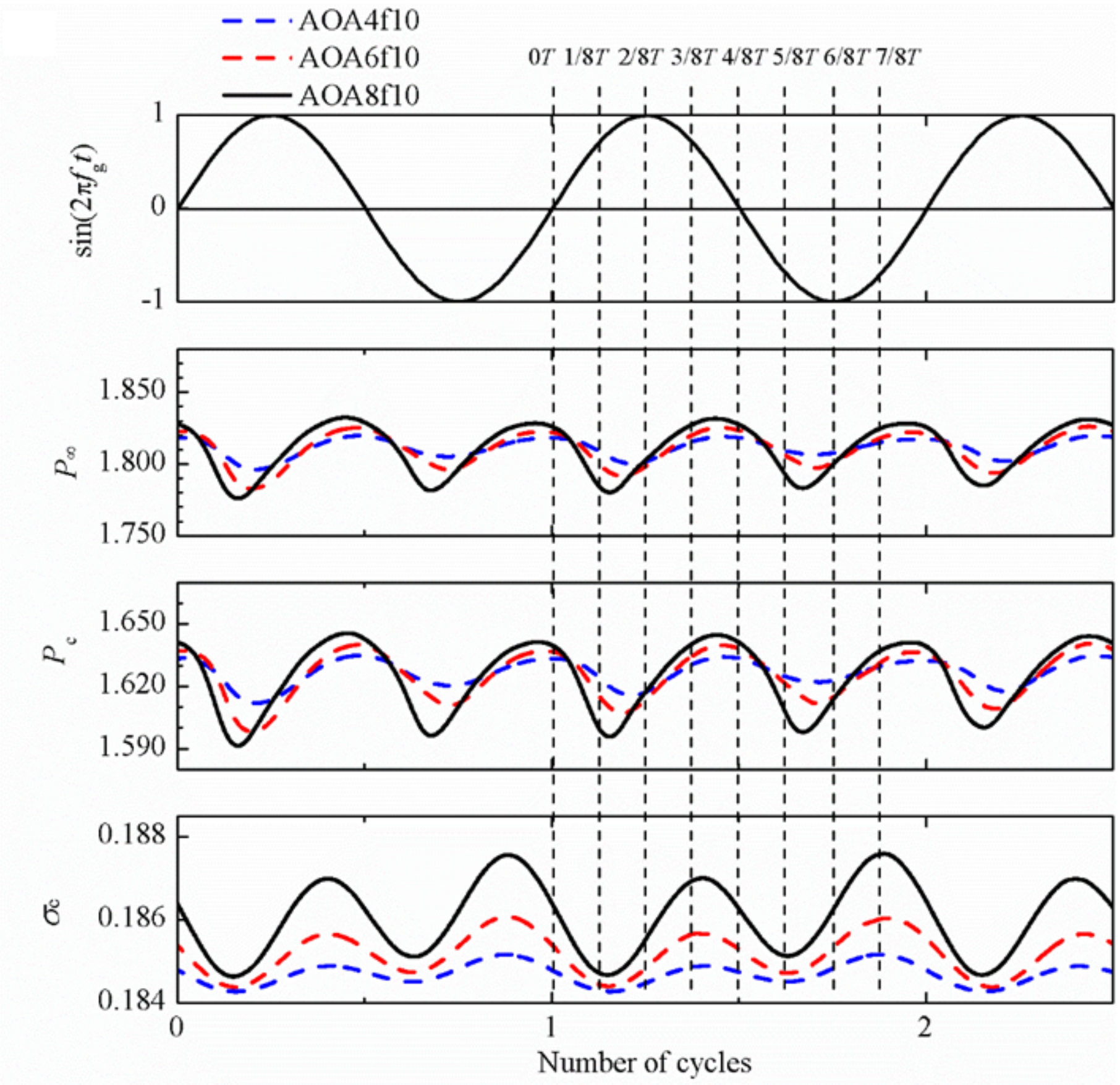

Fig. 14: Effects of the angle of attack ($\gamma_g$) on the non-dimensionalized test section and cavity pressure and cavitation number under the pitching frequency $f_g = 10$ Hz and the angles of attack $\gamma_g = 4°, 6°, 8°$.

Fig. 14 presents the variation of test section pressure and cavity pressure in a gust cycle under the three angles of attack. As it shows, the pressure signals still exhibit periodic variation and a double-peak pattern in a gust cycle, but with a reduced amplitude corresponding to the decrease of the angle of attack. It is worth noting that there is a slight shift of the undulating phase of the pressure signals across different cases of $\gamma_g$ (i.e., instantaneous pressure of high angle of attack displays larger delaying to the phase change of the hydrofoils comparing to smaller angles of attack cases). We attribute such shift to the increasing blockage effect due to the stronger shape deformation for high angle of attack case. In addition to the cavity shape and the pressure signals, the internal flow of the supercavity exhibits similar features at different cases of $\gamma_g$ with increasing reverse flow amplitude and increasing thickness of the boundary layers for high $\gamma_g$ (not shown for brevity).

### 3.5 Effect of the gust frequency

The effect of gust frequency is studied through the simulation under three gust frequencies, i.e., $f_g = 1$ Hz, 5 Hz, 10 Hz, with a fixed angle of attack at $\gamma_g = 8°$ at the same inlet flow speed and ventilation rate. With increasing $f_g$, the gust wavelength $\lambda_g = U_\infty/f_g$ and wave amplitude characterized by maximum vertical velocity of the disturbance ($v_{gmax}$) decrease. When the gust wavelength becomes comparable to the cavity length (corresponding to $f_g = 10$ Hz), the overall cavity centerline exhibits clear wavelike undulation (Fig. 15). Such undulation behavior diminishes significantly with reducing $f_g$. For all three $f_g$, the pressure signals fluctuate periodically with a double-peak in a gust cycle as shown in Fig. 16. However, both test section pressure and cavity pressure have stronger fluctuations with increasing $f_g$ associated with the augmented blockage effect and shape deformation at higher $f_g$. Similar to the case of increasing $\gamma_g$, the pressure signals for $f_g = 10$ Hz shows a slightly larger delay to hydrofoil phase in comparison to the case of 1 Hz. This phase difference is likely to be caused by the larger cavity deformation for $f_g = 10$ Hz causing increased blockage that lowers the advection speed of the incoming flow perturbation in a closed tunnel. In contrast, the cavitation number at lower $f_g$ displays much stronger fluctuation. Such trend can be explained from the comparison of internal flow results across three cases. Under the cases of lowering $f_g$, the cavity has remarkably stronger reverse flow in amplitude and a thicker boundary layer. Correspondingly, although the cavity shape variation diminishes with reducing $f_g$, the cavity exhibits a more prominent closure variation associated with larger cavitation number fluctuation. Such trend of cavity closure variation under different $f_g$ has also been reported in experiment (Shao et al. 2018) and can be explained by the theoretical analysis in Karn et al. (2016) and Karn et al. (2015) which connect dimensionless differential pressure $\Delta\tilde{P}$ introduced in Section 3.2 to the cavitation number.

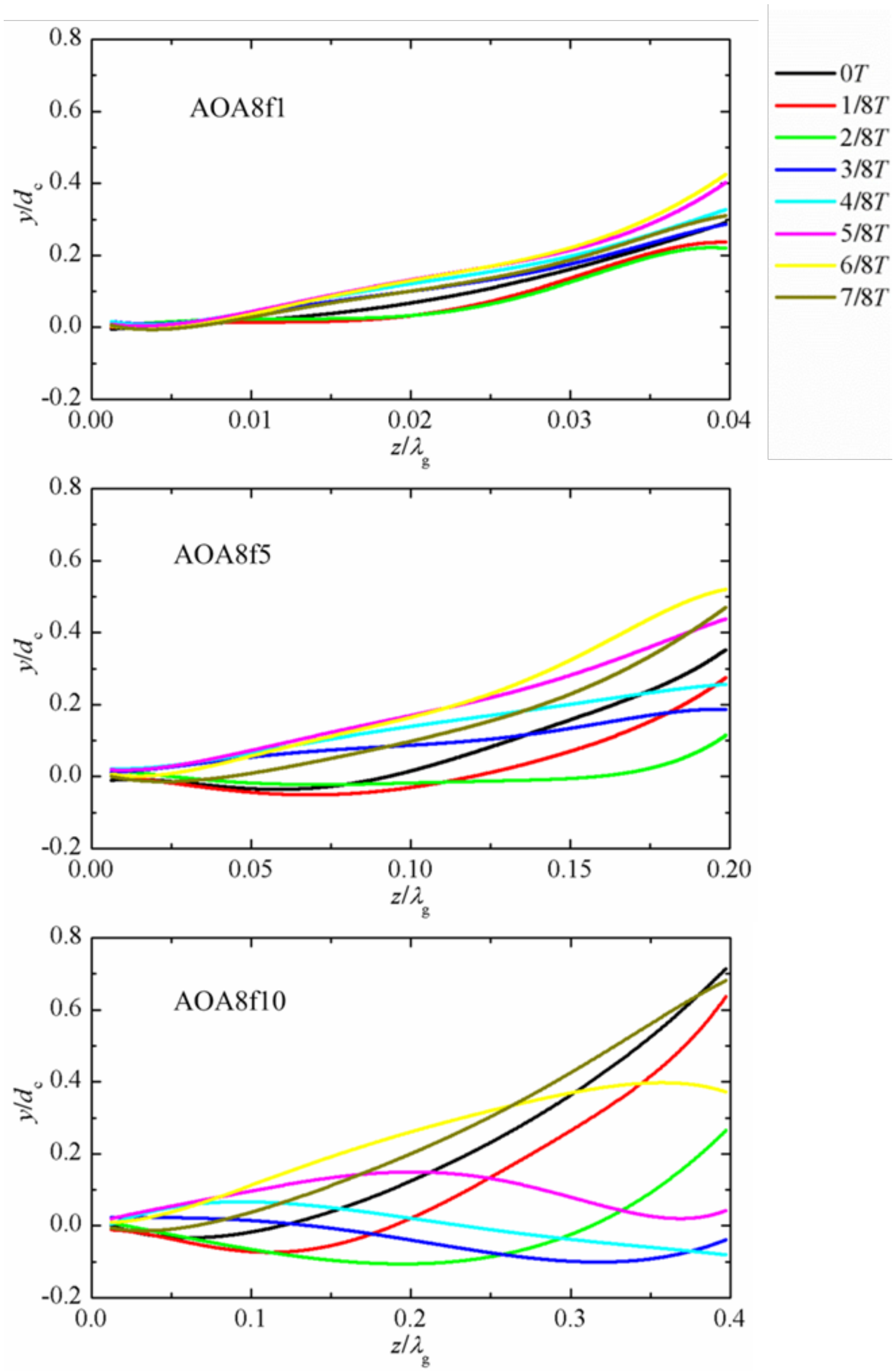

Fig. 15: Effects of the gust frequency ($f_g$) on the variations of supercavity centerlines under the angle of attack $\gamma_g = 8°$ and the gust frequencies $f_g = 1, 5, 10$ Hz.

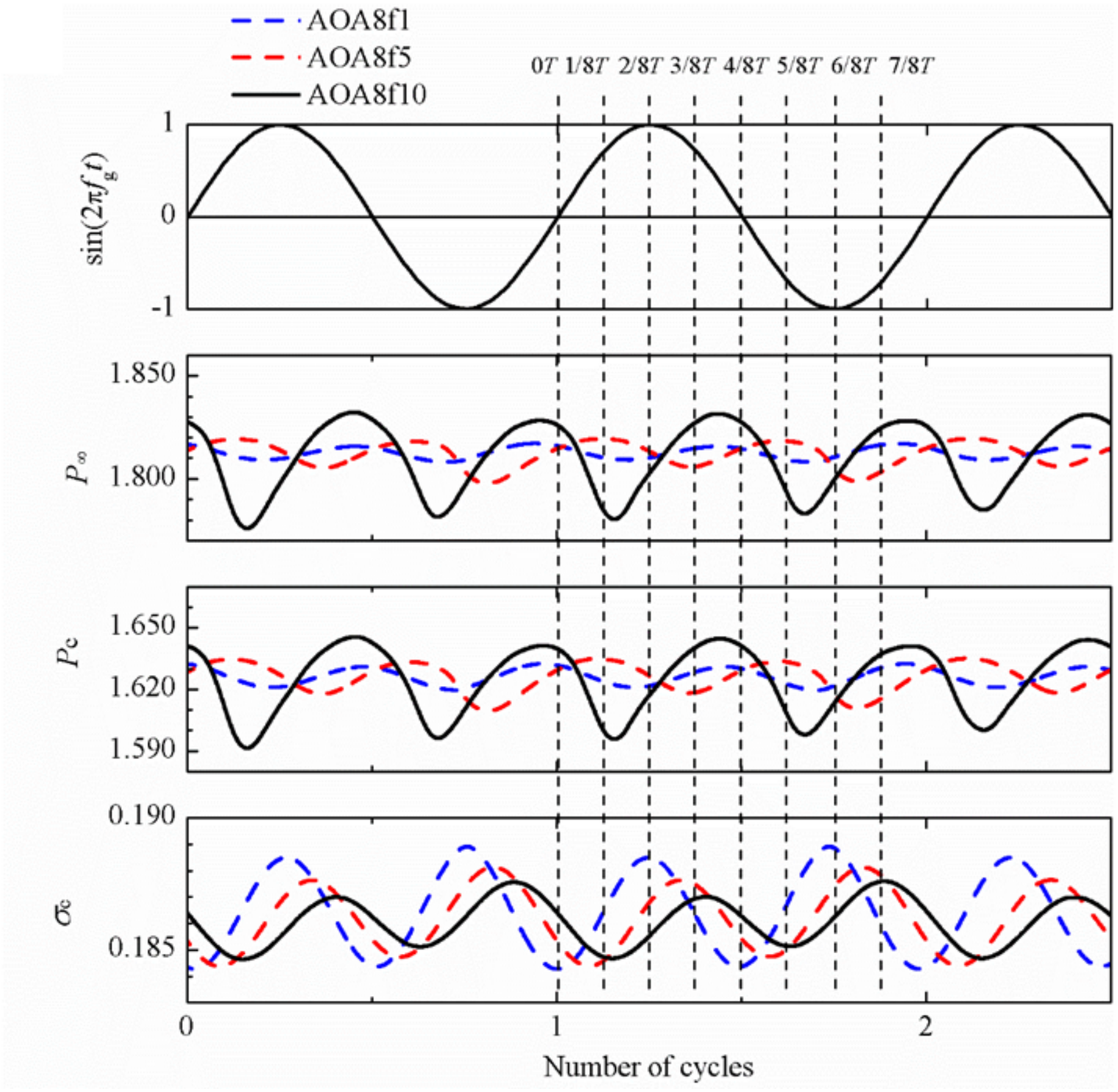

Fig. 16: Effects of the gust frequency ($f_g$) on the non-dimensionalized test section and cavity pressure and cavitation number under the angle of attack $\gamma_g = 8°$ and the gust frequencies $f_g = 1, 5, 10$ Hz.

It is worth noting that the phase of pressure fluctuation for $f_g = 5$ Hz is ahead of the other two cases. We suggest that such difference is related to the phase difference of cavity shape variation under different gust frequencies. Specifically, for $f_g = 5$ Hz, the instant of maximum cavity shape deformation is at $t = 6/8\,T$ while that for $f_g = 10$ Hz takes place at $t = 7/8\,T$ (Fig. 15). Therefore, for $f_g = 5$ Hz, the influence of blockage kicks in ahead that for $f_g = 10$ Hz, leading to the phase difference in their pressure signals. However, for $f_g = 1$ Hz, the overall shape variation of cavity is negligible comparing to the other cases, and thus does not alter the phases of the pressure. In comparison to the experimental results under the same flow conditions and cavitator model reported in Shao et al. (2018), the simulation does not capture the change of supercavity mode as it transitions from wavelike undulation to cavity pulsating with reduction of $f_g$. Specifically, the experiment shows drastic change of cavity lengths due to the shed-off of gas pocket at the rear of the cavity. Such instability has not been captured in our simulation due to the limitation of the numerical approach. This limitation is discussed in detail in Section 4. Nevertheless, the simulation can still provide an overall degree and phase of wavelike undulation occurring on cavity surface under different unsteady conditions.

## 4. Summary and Discussions

The ventilated supercavitation from a forward facing cavitator under periodic gust flow is numerically studied by using the Eulerian multiphase model with a free surface model. The flow unsteadiness is simulated by adding a harmonic varying vertical velocity term matching that produced from a pair of flapping hydrofoils in the experiments (Lee et al. 2013, Shao et al. 2018). The numerical method is first validated through the experiments under specific steady and unsteady flow conditions. It has been shown that the simulation can capture the degree of cavity shape fluctuation and internal pressure variation in a gust cycle. Specifically, the cavity centerline shows periodic wavelike undulation with a maximum amplitude matching that of the incoming flow perturbation. The cavity internal pressure also fluctuates periodically, causing the corresponding change of difference between internal and external pressure across the closure that leads to the closure mode change in a gust cycle. In addition, the simulation captures the variation of cavity internal flow, particularly the development internal flow boundary layer along the cavitator mounting strut, upon the incoming flow perturbation, correlating with cavity deformation and closure mode variation. With increasing angle of attack, the cavity exhibits augmented wavelike undulation and pressure fluctuation. As the wavelength of the flow perturbation approaches the cavity length with increasing gust frequency, the cavity experiences stronger wavelike undulation and internal pressure fluctuation but reduced cavitation number variation.

The numerical simulation provides detailed information of cavity internal flow and pressure distribution which are very difficult to be measured even with the most recent experimental investigation employing sophisticated imaging techniques (Wu et al. 2019). Specifically, the simulation result captures the development of internal flow throughout the entire cavity, including the region near the closure region which is not accessible in Wu et al. (2019). Further, in comparison to backward-facing cavitator used in Wu et al. (2019), the simulation is conducted on a forward-facing cavitator and shows the change of internal flow structures due to the influence of the mounting strut, such as the enhanced reverse flow due to the formation of the boundary layer along the mounting strut. The internal pressure distribution provided from the simulation allows us to directly assess the variation of pressure difference across the closure region of the cavity. The result not only provides support to the framework proposed by Karn et al. (2015) for explaining the supercavity closure change under different steady flow conditions, but also suggests the applicability of such framework for predicting cavity closure variation in unsteady flows.

Based on the simulation under different angles of attack and gust frequencies, we have shown that the phases of cavity deformation, particularly, the location and the time instant corresponding to the maximum vertical displacement of the cavity, vary according to the wavelength and wave amplitude of unsteady flow. Such variation needs to be considered in the ventilation design and the control of the supercavitating devices. Additionally, the simulation shows that the varying degree of cavity deformation can cause phase shift of pressure signals under different unsteady conditions in a closed water tunnel. This observation further underscores the importance of understanding the effect from flow facilities on supercavity behavior as discussed in Shao et al. (2017).

Finally, it is noteworthy that the present simulation is unable to capture the cavity length change due to shed-off of gas pocket near its closure region (i.e., pulsation of the cavity) which occurs under the cases of AOA8f1 and AOA8f5 in the experiments (Shao et al. 2018). As discussed in (Shao et al. 2018), the transition from the wavelike undulation to the cavity pulsation is a result of the bubble breakup near cavity closure induced by the instability of the gas-liquid interface. Therefore, to fully capture cavity behaviors like pulsation, future numerical simulation should consider detailed bubble-liquid interaction such as bubble induced turbulence, lift force, and drift velocity of bubbles which are not explicitly included in the present simulation. As an example, in the modeling of a bubble column, Selma et al. (2010) considered additional terms like lift force on individual bubbles, bubble drifting due to turbulence, and virtual mass force in the exchange of momentum on gas-liquid interface. A standard $k-\varepsilon$ model with bubble induced turbulence was used for turbulence modeling. The simulation results have shown to be able to resolve the vortex structures developed on the gas-liquid interface due to flow instability. Large eddy simulation (LES) with locally adaptive eddy viscosity terms provides another approach to capture the cavity pulsation in unsteady flows. It was recently employed to capture the pulsation of a hydrofoil cavitation originated from flow instability on the vapor-water interface (Ji et al. 2015). Such high-fidelity simulation should be applied to the study of ventilated supercavitation in unsteady flows, which can provide more physical insights into its behaviors.

## Acknowledgment

Renfang Huang is supported by the State Scholarship Fund from China Scholarship Council. We also acknowledge the support from the Office of Naval Research (Program Manager, Dr.

Thomas Fu) under grant No. N000141612755 and the start-up funding received by Prof. Jiarong Hong from University of Minnesota.

## Nomenclature

| | |
|---|---|
| $\sigma$ | ventilated cavitation number |
| $Fr$ | Froude number |
| $C_Q$ | air entrainment coefficient |
| $p_\infty$ | ambient pressure upstream of the cavitator |
| $p_c$ | cavity pressure |
| $\rho_w$ | water density, =997 kg/m$^3$ |
| $U_\infty$ | upstream incoming flow velocity |
| g | gravitational acceleration |
| $d_c$ | cavitator diameter |
| $\dot{Q}$ | volumetric ventilation rate |
| $\alpha_k$ | volume fraction of phase *k* |
| $\rho_k$ | density of phase *k* |
| $\mathbf{u}_k$ | mean velocity vector |
| $\rho_a$ | air density, =1.185 kg/m$^3$ |
| $p$ | pressure |
| $\mu_k$ | molecular dynamic viscosity of phase $k$ |
| $\mu_{t,k}$ | turbulent eddy viscosity of phase $k$ |
| $\mathbf{S}_{k,\mathrm{buoy}}$ | momentum source due to buoyancy forces |
| $\mathbf{M}_k$ | total interfacial drag forces |
| $C_\mathrm{D}$ | non-dimensional drag coefficient, =0.44 |
| $\rho_\mathrm{m}$ | density of the mixture |
| $A_\mathrm{m}$ | interfacial area per unit volume |
| $v_\mathrm{g}(t)$ | vertical velocity term, $v_\mathrm{g}(t) = v_\mathrm{gmax}\sin(2\pi f_\mathrm{g} t)$ |
| $v_\mathrm{gmax}$ | maximum vertical velocity (amplitude) |
| $\gamma_\mathrm{g}$ | angle of attack of two hydrofoils |
| $f_\mathrm{g}$ | frequency of periodic flows (wavelength) |
| $D_\mathrm{max}$ | maximum diameter of the supercavity |
| $L_{1/2}$ | half-length of the supercavity |
| $\lambda_\mathrm{g}$ | gust wavelength, $\lambda_\mathrm{g} = U_\infty / f_\mathrm{g}$ |
| $\Omega_\mathrm{z}$ | vorticity component |
| $C_\mathrm{p}$ | pressure coefficient, $= 2(p - p_\infty)/(\rho_\mathrm{w} U_\infty^2)$ |
| $\Delta\tilde{P}$ | dimensionless pressure difference |
| $\tilde{P}_\mathrm{in}, \tilde{P}_\mathrm{out}$ | dimensionless static pressure inside and just outside the cavity at the closure |

$\tilde{P}_\infty$ dimensionless test section pressure

$\tilde{P}_c$ dimensionless cavity pressure